\documentclass[onecolumn,showpacs,preprintnumbers,showkeys]{revtex4}%
\usepackage[format=default]{caption}
\usepackage{amsfonts}
\usepackage{amsmath}
\usepackage{graphicx}
\usepackage{dcolumn}
\usepackage{bm}
\usepackage{amssymb}
\usepackage{latexsym}%
\begin{document}
\title{Vacuum Properties of Mesons in a Linear Sigma Model with Vector Mesons and
Global Chiral Invariance}
\author{Denis Parganlija$^{\text{(a)}}$, Francesco Giacosa$^{\text{(a)}}$, and Dirk
H.\ Rischke$^{\text{(a,b)}}$}
\affiliation{$^{\text{(a)}}$Institute for Theoretical Physics, Johann Wolfgang Goethe
University, Max-von-Laue-Str.\ 1, D--60438 Frankfurt am Main, Germany}
\affiliation{$^{\text{(b)}}$Frankfurt Institute for Advanced Studies, Ruth-Moufang-Str.\ 1,
D--60438 Frankfurt am Main, Germany}

\begin{abstract}
We present a two-flavour linear sigma model with global chiral symmetry and
vector and axial-vector mesons. We calculate $\pi\pi$ scattering lengths and
the decay widths of scalar, vector, and axial-vector mesons. It is
demonstrated that vector and axial-vector meson degrees of freedom play an
important role in these low-energy processes and that a reasonable theoretical
description requires globally chirally invariant terms other than the vector
meson mass term. An important question for meson vacuum phenomenology is the
quark content of the physical scalar $f_{0}(600)$ and $a_{0}(980)$ mesons. We
investigate this question by assigning the quark-antiquark $\sigma$ and
$a_{0}$ states of our model with these physical mesons. We show via a detailed
comparison with experimental data that this scenario can describe all vacuum
properties studied here except for the decay width of the $\sigma$, which
turns out to be too small. We also study the alternative assignment
$f_{0}(1370)$ and $a_{0}(1450)$ for the scalar mesons. In this case the decay
width agrees with the experimental value, but the $\pi\pi$ scattering length
$a_{0}^{0}$ is too small. This indicates the necessity to extend our model by
additional scalar degrees of freedom.

\end{abstract}

\pacs{12.39.Fe, 13.75.Lb, 13.20.Jf}
\keywords{chiral Lagrangians, global invariance, pion-pion scattering, decay widths,
scalar mesons.}

\maketitle

\bigskip

\section{Introduction}

The fundamental theory of strong interactions, Quantum Chromodynamics (QCD),
possesses an exact $SU(3)_{c}$ local gauge symmetry (the color symmetry) and
an approximate global $U(N_{f})_{R}\times U(N_{f})_{L}$ symmetry for $N_{f}$
massless quark flavours (the chiral symmetry). For sufficiently low
temperature and density quarks and gluons are confined into colorless hadrons
(i.e., $SU(3)_{c}$ invariant configurations). Thus, it is the chiral symmetry
which predominantly determines hadronic interactions in the low-energy region.

Effective field theories which contain hadrons as degrees of freedom rather
than quarks and gluons have been developed along two lines which differ in the
way in which chiral symmetry is realized: linear \cite{gellmanlevy} and
non-linear \cite{weinberg}. In the non-linear realization, the
so-called non-linear sigma model, the scalar states are integrated
out, leaving the pseudoscalar states as the only degrees of the
freedom. On the other hand, in the linear representation of the
symmetry, the so-called linear sigma model, both
the scalar and pseudoscalar degrees of freedom are present.

In this work, we consider the linear representation of chiral symmetry. An
exactly linearly realized chiral symmetry implies that the QCD eigenstates
come in degenerate pairs, the so-called chiral partners. Chiral partners have
the same quantum numbers with the exception of parity and G-parity -- for
example, the scalar states sigma and pion and the vector states $\rho$ and
$a_{1}$, respectively, are chiral partners. Experimental data in vacuum and at
sufficiently low temperatures and densities of matter, however, show that the
mass degeneracy is lifted, because the chiral $U(N_{f})_{R}\times U(N_{f})_{L}
\equiv U(1)_{V}\times U(1)_{A}\times SU(N_{f})_{V}\times SU(N_{f})_{A}$
symmetry is broken in two ways: explicitly and spontaneously.

Due to the $U(1)_{A}$ anomaly \cite{Hooft}, the $U(N_{f})_{R}\times
U(N_{f})_{L}$ symmetry is broken explicitly by quantum effects to
$U(1)_{V}\times SU(N_{f})_{V}\times SU(N_{f})_{A}$. In the case of small but
nonzero degenerate quark masses, the latter is explicitly broken to
$U(N_{f})_{V}$. If the quark masses are not degenerate, the $U(N_{f})_{V}$
symmetry is furthermore explicitly broken to $U(1)_{V}$, corresponding to
baryon number conservation. QCD also possesses discrete symmetries such as the
charge conjugation (\textit{C}), parity (\textit{P}) and time reversal
(\textit{T}) symmetry (\textit{CPT}), which are to a very good precision separately
conserved by strong interactions. This fact offers further constraints in the
construction of effective models of QCD. [A review of a possible, although
small, \textit{CP} violation in strong interactions may be found e.g.\
in Ref.\ \cite{Peccei}.]

In addition to the explicit breaking of axial symmetry $SU(N_{f})_{A}$ due to
nonzero quark masses, the latter symmetry is also spontaneously broken in
vacuum by the non-vanishing expectation value of the quark condensate:
$\langle\bar{q}q\rangle=\langle\bar{q}_{R}q_{L}+\bar{q}_{L} q_{R}\rangle\neq
0$\ \cite{SSB}. This symmetry breaking mechanism leads to the emergence of
$N_{f}^{2}-1$ pseudoscalar Goldstone bosons, as well as of massive scalar
states representing the chiral partners of the Goldstone bosons. For $N_{f}%
=2$, the three lightest mesonic states, the pions, are identified with these
Goldstone bosons of QCD. Their non-vanishing mass arises due to the explicit
breaking of the chiral symmetry, rendering them pseudo-Goldstone bosons.

In this paper we study an $N_{f}=2$ linear sigma model which contains scalar
($\sigma$, $\vec{a}_{0}$) and pseudoscalar ($\eta$, $\vec{\pi}$), and in
addition also vector ($\omega$, $\vec{\rho}$) and axial-vector ($f_{1}$,
$\vec{a}_{1}$) degrees of freedom. Usually, such models are constructed under
the requirement of local chiral invariance $U(N_{f})_{R}\times U(N_{f})_{L}$,
with the exception of the vector meson mass term which renders the local
symmetry a global one \cite{GG,KR}. In a slight abuse of terminology, we will
refer to these models as locally chirally invariant models in the following. A
study of the QCD phase transition and its critical temperature $T_{c}$ within such
a model can be found e.g.\ in Ref.\ \cite{RS}.

However, as shown in Refs.\ \cite{GG,KR,Meissner,Lissabon,Mainz}, the locally
invariant linear sigma model fails to simultaneously describe meson decay
widths and pion-pion scattering lengths in vacuum. As outlined in
Ref.\ \cite{Lissabon}, there are at least two ways to solve this issue. One
way is to utilize a model in which the (up to the vector meson mass term)
local invariance of the theory is retained while higher-order terms are added
to the Lagrangian \cite{GG,KR,Meissner}. The second way which is pursued here
is the following: we construct a linear sigma model with global chiral
invariance containing all terms up to naive scaling dimension four \cite{UBW}.
The global invariance allows for additional terms to appear in our Lagrangian
in comparison to the locally invariant case presented e.g.\ in Ref.\ \cite{RS}%
. We remark that, introducing a dilaton field, one can argue
\cite{Susanna,dynrec} that chirally invariant terms of higher order than
scaling dimension four should be absent.

In Ref.\ \cite{Mainz}, we have presented a first study of meson decays and
pion-pion scattering lengths in vacuum in the framework of the globally
invariant linear sigma model. We have distinguished two different assignments
for the scalar fields $\sigma=\frac{1}{\sqrt{2}}(\bar{u}u+\bar{d}d)$ and
$a_{0}^{0}=\frac{1}{\sqrt{2}}(\bar{u}u-\bar{d}d)$: (\textit{i}) they may be
identified with $f_{0}(600)$ and $a_{0}(980)$ which are members of a nonet
that in addition consists of $f_{0}(980)$ and $\kappa(800)$; (\textit{ii})
they may be identified with $f_{0}(1370)$ and $a_{0}(1450)$ which are members
of a decuplet that in addition consists of $f_{0}(1500)$, $f_{0}(1710)$, and
$K_{0}(1430)$, where the additional scalar-isoscalar state emerges from the
admixture of a glueball field \cite{refs1}. In the following, we will refer to
assignment (\textit{i}) as Scenario I, and to assignment (\textit{ii}) as
Scenario II. In the latter, scalar mesons below 1 GeV are not (predominantly)
quark-antiquark states. Their spectroscopic wave functions might contain a
dominant tetraquark or mesonic molecular contribution \cite{refs2}. The
correct assignment of the scalar quark-antiquark fields of the model to
physical resonances is not only important as a contribution to the ongoing
debate about the nature of these resonances, but it is also vital for a
study of the properties of hadrons at nonzero temperature and density,
where the chiral partner of the pion plays a crucial role \cite{Heinz}.

It is important to stress that the theoretical $\sigma$ and $a_{0}$ fields
entering the linear sigma model describe pure quark-antiquark states, just as
all the other fields ($\eta$, $\vec{\pi}$, $\omega$, $\vec{\rho}$, $f_{1}$,
$\vec{a}_{1}$). This property can be easily proven by using well-known
large-$N_{c}$ results \cite{largenc}: the mass and the decay widths of both
$\sigma$ and $a_{0}$ fields scale in the model as $N_{c}^{0}$ and
$N_{c}^{-1}$, respectively.

In this paper we first investigate the consequences of Scenario I on various
decay widths and pion-pion scattering lengths. This assignment is disfavored
because a consistent description of all experimental data cannot be achieved.
To reach this conclusion, vector and axial-vector degrees of freedoms play an
important role. On the one hand their decays (such as $\rho\rightarrow\pi\pi$
and $a_{1}\rightarrow\pi\gamma$) and the role of the $\rho$ meson in $\pi\pi$
scattering provide strong constraints, on the other hand they affect,
indirectly but sizably, some decay channels, such as $\sigma\rightarrow\pi\pi
$. We then present a study of Scenario II. Although the latter is not yet
conclusive because additional scalar fields (glueball, tetraquark) are not yet
taken into account, our preliminary results for the decay widths (albeit not
for the scattering length $a_{0}^{0}$) are consistent with the data.

The paper is organized as follows: in Sec.\ II we present the Lagrangian of
our model and discuss the parameters which are known to very good precision
and thus do not enter the fit of the decay widths and the scattering lengths.
In Sec.\ III we present the formulas for the decay widths and the pion-pion
scattering lengths which will be used to fit the remaining parameters and to
compare the results to experimental data. This fit and comparison are
discussed in Sec.\ IV, both for Scenario I and Scenario II. In Sec.\ V we
summarize our results in the conclusions and give an outlook to future
work. In the Appendix, we show the
explicit form of our Lagrangian in terms of the meson fields.

\section{The Linear Sigma Model with Global Chiral Symmetry}

\subsection{The Lagrangian}

The Lagrangian of the globally invariant linear sigma model with
$U(2)_{R}\times U(2)_{L}$ symmetry for $N_{f}=2$ reads
\cite{GG,KR,Mainz,Boguta}:
\begin{align}
\mathcal{L}  &  =\mathrm{Tr}[(D^{\mu}\Phi)^{\dagger}(D^{\mu}\Phi)]-m_{0}^{2}
\mathrm{Tr}(\Phi^{\dagger}\Phi)-\lambda_{1}[\mathrm{Tr}(\Phi^{\dagger}%
\Phi)]^{2} -\lambda_{2}\mathrm{Tr}(\Phi^{\dagger}\Phi)^{2}\nonumber\\
&  -\frac{1}{4}\mathrm{Tr}[(L^{\mu\nu})^{2}+(R^{\mu\nu})^{2}]+\frac{m_{1}^{2}%
}{2} \mathrm{Tr}[(L^{\mu})^{2}+(R^{\mu})^{2}]+\mathrm{Tr}[H(\Phi+\Phi
^{\dagger})]\nonumber\\
&  +c(\det\Phi+\det\Phi^{\dagger})-2ig_{2}(\mathrm{Tr}\{L_{\mu\nu}[L^{\mu
},L^{\nu}]\} +\mathrm{Tr}\{R_{\mu\nu}[R^{\mu},R^{\nu}]\})\nonumber\\
&  -2g_{3}\left[  \mathrm{Tr}\left(  \left\{  \partial_{\mu}L_{\nu}-ieA_{\mu
}[t^{3},L_{\nu}] +\partial_{\nu}L_{\mu}-ieA_{\nu}[t^{3},L_{\mu}]\right\}
\{L^{\mu},L^{\nu}\}\right)  \right. \nonumber\\
&  \hspace*{0.65cm} +\left.  \mathrm{Tr} \left(  \left\{  \partial_{\mu}%
R_{\nu}-ieA_{\mu}[t^{3},R_{\nu}] +\partial_{\nu}R_{\mu}-ieA_{\nu}[t^{3}%
,R_{\mu}]\right\}  \{R^{\mu},R^{\nu}\}\right)  \right] \nonumber\\
&  +\frac{h_{1}}{2}\mathrm{Tr}(\Phi^{\dagger}\Phi)\mathrm{Tr}[(L^{\mu})^{2}
+(R^{\mu})^{2}]+h_{2}\mathrm{Tr}[(\Phi R^{\mu})^{2}+(L^{\mu}\Phi)^{2}]
+2h_{3}\mathrm{Tr}(\Phi R_{\mu}\Phi^{\dagger}L^{\mu}).\nonumber\\
&  +g_{4}\left\{  \mathrm{Tr}\left[  L^{\mu}L^{\nu}L_{\mu}L_{\nu}\right]
+\mathrm{Tr}\left[  R^{\mu}R^{\nu}R_{\mu}R_{\nu}\right]  \right\}
+g_{5}\left\{  \mathrm{Tr}\left[  L^{\mu}L_{\mu}L^{\nu}L_{\nu}\right]
+\mathrm{Tr}\left[  R^{\mu}R_{\mu}R^{\nu}R_{\nu}\right]  \right\} \nonumber\\
&  +g_{6}\mathrm{Tr}\left[  R^{\mu}R_{\mu}\right]  \,\mathrm{Tr}\left[
L^{\nu}L_{\nu}\right]  +g_{7}\left\{  \mathrm{Tr}[L^{\mu}L_{\mu}]
\,\mathrm{Tr}[L^{\nu}L_{\nu}]+\mathrm{Tr}[R^{\mu}R_{\mu}]\,\mathrm{Tr}
[R^{\nu}R_{\nu}]\right\}  \;. \label{Lagrangian}%
\end{align}
Note that the locally chirally invariant linear sigma model emerges from the
globally invariant Lagrangian (\ref{Lagrangian}) by setting $h_{1}=h_{2}%
=h_{3}=g_{3}=0,$ $g_{2}=g_{4}=g_{5}=g_{6}=g_{7} \equiv g$.

In Eq.\ (\ref{Lagrangian}),
\begin{equation}
\Phi=(\sigma+i\eta_{N})\,t^{0}+(\vec{a}_{0}+i\vec{\pi})\cdot\vec{t}
\label{scalars}%
\end{equation}
contains scalar and pseudoscalar mesons, where $t^{0}$, $\vec{t}$ are the
generators of $U(2)$ in the fundamental representation and $\eta_{N}$ denotes
the non-strange content of the $\eta$ meson. Vector and axial-vector mesons
are contained in the left-handed and right-handed vector fields:
\begin{subequations}
\begin{align}
L^{\mu}  &  = (\omega^{\mu}+f_{1}^{\mu})\,t^{0}+(\vec{\rho}^{\mu}+\vec{a}%
_{1}^{\mu}) \cdot\vec{t} \;, \label{vectors1}\\
R^{\mu}  &  = (\omega^{\mu}-f_{1}^{\mu})\,t^{0}+(\vec{\rho}^{\mu}-\vec{a}%
_{1}^{\mu}) \cdot\vec{t}\; , \label{vectors}%
\end{align}
\end{subequations}
respectively. The covariant derivative
\begin{equation}
D^{\mu}\Phi=\partial^{\mu}\Phi-ig_{1}(L^{\mu}\Phi-\Phi R^{\mu})-ieA^{\mu
}[t^{3},\Phi]
\end{equation}
couples scalar and pseudoscalar degrees of freedom to vector and axial-vector
ones as well as to the electromagnetic field $A^{\mu}$. Note that local chiral
invariance requires $g_{1} \equiv g$. The left-handed and right-handed field
strength tensors,
\begin{subequations}
\begin{align}
L^{\mu\nu}  &  = \partial^{\mu}L^{\nu}-ieA^{\mu}[t^{3},L^{\nu}] -\left\{
\partial^{\nu}L^{\mu}-ieA^{\nu}[t^{3},L^{\mu}]\right\}  \;,\\
R^{\mu\nu}  &  = \partial^{\mu}R^{\nu}-ieA^{\mu}[t^{3},R^{\nu}] -\left\{
\partial^{\nu}R^{\mu}-ieA^{\nu}[t^{3},R^{\mu}]\right\}  \;,
\end{align}
respectively, couple vector and axial-vector mesons to the electromagnetic
field $A^{\mu}$. Explicit breaking of the global symmetry is described by the
term Tr$[H(\Phi+\Phi^{\dagger})]\equiv h_{0}\sigma$ $(h_{0}=const.)$. The
chiral anomaly is described by the term $c\,(\det\Phi+\det\Phi^{\dagger})$
\cite{Hooft}. The model has been extended to include the nucleon field and its
putative chiral partner; for details, see Refs.\ \cite{Susanna,Susanna-Mainz}.

In the pseudoscalar and (axial-)vector sectors the identification of mesons
with particles listed in Ref.\ \cite{PDG} is straightforward, as already
indicated in Eqs.\ (\ref{scalars}) and
(\ref{vectors1})-(\ref{vectors}): the fields $\vec{\pi}$
and $\eta_{N}$ correspond to the pion and the $SU(2)$ counterpart of the
$\eta$ meson, $\eta_{N}\equiv(\overline{u}u+\overline{d}d)/\sqrt{2}$, with a
mass of about $700$ MeV. This value can be obtained by "unmixing" the
physical $\eta$ and $\eta^{\prime}$ mesons, which also contain $\overline{s}s$
contributions. The fields $\omega^{\mu}$ and $\vec{\rho}^{\mu}$ represent the
$\omega(782)$ and $\rho(770)$ vector mesons, respectively, while the fields
$f_{1}^{\mu}$ and $\vec{a_{1}}^{\mu}$ represent the $f_{1}(1285)$ and
$a_{1}(1260)$ axial-vector mesons, respectively. (In principle, the physical
$\omega$ and $f_{1}$ states also contain $\overline{s}s$ contributions,
however their admixture is negligibly small.) Unfortunately, the
identification of the $\sigma$ and $\vec{a}_{0}$ fields is controversial, the
possibilities being the pairs $\{f_{0}(600),a_{0}(980)\}$ and $\{f_{0}%
(1370),a_{0}(1450)\}$. As mentioned in the Introduction, we 
will refer to these two assignments as Scenarios I
and II, respectively. We discuss the implications of these two scenarios in
the following.

One may raise the question whether vector meson dominance (VMD) is still
respected in the globally invariant linear sigma model (\ref{Lagrangian}). As
outlined in Ref.\ \cite{OCPTW}, there are two ways to realize VMD in a linear
sigma model. The standard version of VMD was introduced by Sakurai
\cite{Sakurai} and considers vector mesons as Yang-Mills gauge fields
\cite{YM}. The gauge symmetry is explicitly broken by the vector meson masses.
Another realisation of VMD was first explored by Lurie \cite{Lurie} whose
theory contained a Lagrangian which was globally invariant. It is interesting
to note that Lurie's Lagrangian contained direct couplings of the photon to
pions and $\rho$ mesons, as well as a $\rho$-$\pi$ coupling. It was shown in
Ref.\ \cite{OCPTW} that the two representations of VMD are equivalent if the
$\rho$-$\pi$ coupling $g_{\rho\pi\pi}$ equals the photon-$\rho$ coupling
$g_{\rho}$ (the so-called "universal limit"). It was also shown that, if the
underlying theory is globally invariant, the pion form factor at threshold
$F_{\pi}(q^{2}=0)=1$ for \emph{any\/} value of the above mentioned couplings.
On the other hand, in Sakurai's theory $F_{\pi}(q^{2}=0)\neq1$ unless one
demands $g_{\rho\pi\pi}\overset{!}{=}g_{\rho}$, or other parameters are
adjusted in such a way that $F_{\pi}(q^{2}=0)=1$. In other words, for
\emph{any\/} globally invariant model, and thus also for ours, one has the
liberty of choosing different values for the photon-$\rho$ and $\rho$-$\pi$
couplings, without violating VMD.

\subsection{Tree-Level Masses}

The Lagrangian (\ref{Lagrangian}) contains 16 parameters. However, the
parameters $g_{k}$ with $k=3,$ ..., $7$ are not relevant for the results
presented here so that the number of undetermined parameters decreases to
eleven:
\end{subequations}
\begin{equation}
m_{0},\text{ }\lambda_{1},\text{ }\lambda_{2},\text{ }m_{1},\text{ }
g_{1},\text{ }g_{2},\text{ }c,\text{ }h_{0},\text{ }h_{1}, \text{ }h_{2},
\text{ }h_{3} \;. \label{param}%
\end{equation}
The squared tree-level masses of the mesons in our model contain a
contribution arising from spontaneous symmetry breaking, proportional to
$\phi^{2}$. The value $\phi$ is the vacuum expectation value of the $\sigma$
field and coincides with the minimum of the potential that follows from
Eq.\ (\ref{Lagrangian}). The $\sigma$ field is the only field with the quantum
numbers of the vacuum, $J^{PC}=0^{++}$, i.e., the condensation of which does
not lead to the breaking of parity, charge conjugation, and Lorentz
invariance. The potential for the $\sigma$ field reads explicitly
\begin{equation}
V(\sigma) =\frac{1}{2}(m_{0}^{2}-c)\sigma^{2}+\frac{1}{4}\left(  \lambda
_{1}+\frac{\lambda_{2}}{2}\right)  \sigma^{4}-h_{0}\sigma\;,
\end{equation}
and its minimum is determined by
\begin{equation}
0 =\left(  \frac{\mathrm{d}V}{\mathrm{d}\sigma}\right)  _{\sigma=\phi}
=\left[  m_{0}^{2}-c+\left(  \lambda_{1}+\frac{\lambda_{2}}{2}\right)
\phi^{2}\right]  \phi-h_{0}\; \text{.} \label{minimum}%
\end{equation}
Spontaneous symmetry breaking corresponds to the case when the potential
$V(\phi)$ assumes its minimum for a nonvanishing value $\sigma=\phi\neq0$. In
order to determine the fluctuation of the $\sigma$ field around the new
vacuum, one shifts it by its vacuum expectation value $\phi\neq0$,
$\sigma\rightarrow\sigma+\phi$. The shift leads also to $\eta_{N}$-$f_{1}$ and
$\vec{\pi}$-$\vec{a}_{1}$ mixing terms and thus to
non-diagonal elements in the scattering matrix. These terms are removed from
the Lagrangian by shifting the $f_{1}$ and $\vec{a}_{1}$ fields as
follows:
\begin{equation}
f_{1}^{\mu}\rightarrow f_{1}^{\mu}+Zw\partial^{\mu}\eta_{N}\;\text{,}\;\;
\vec{a}_{1}^{\mu}\rightarrow\vec{a}_{1}^{\mu}
+Zw\partial^{\mu}\vec{\pi}\;, \;\; \eta_{N}\rightarrow Z\eta_{N}\;
, \;\; \vec{\pi}\rightarrow Z\vec{\pi}\;, \label{shifts}%
\end{equation}
where we defined the quantities
\begin{equation}
w:=\frac{g_{1}\phi}{m_{a_{1}}^{2}}\; , \;\;\; Z:=\left(  1- \frac{g_{1}%
^{2}\phi^{2}}{m_{a_{1}}^{2}}\right)  ^{-1/2}\;\text{.} \label{Z}%
\end{equation}

Note that the field renormalisation of $\eta_{N}$ and $\vec{\pi}$
guarantees the canonical normalization of the kinetic terms. This is necessary
in order to interpret the Fourier components of the properly normalized
one-meson states as creation or annihilation operators \cite{GG}. Note also
that the $\rho$ and $\omega$ masses as well as the $f_{1}$ and $a_{1}$ masses
are degenerate in the globally as well as in the locally invariant model. Once
the shift $\sigma\rightarrow\sigma+\phi$ and the transformations
(\ref{shifts}) have been performed, the mass terms of the mesons in the
Lagrangian (\ref{Lagrangian}) read:
\begin{align}
m_{\sigma}^{2}  &  =m_{0}^{2}-c+3\left(  \lambda_{1}+\frac{\lambda_{2}}%
{2}\right)  \phi^{2}\;,\label{sigma}\\
m_{\eta_{N}}^{2}  &  =Z^{2}\left[  m_{0}^{2}+c+\left(  \lambda_{1}
+\frac{\lambda_{2}}{2}\right)  \phi^{2}\right]  = m_{\pi}^{2}+2cZ^{2}%
\;,\label{eta}\\
m_{a_{0}}^{2}  &  =m_{0}^{2}+c+\left(  \lambda_{1}+3\frac{\lambda_{2}}
{2}\right)  \phi^{2}\;,\label{a0}\\
m_{\pi}^{2}  &  =Z^{2}\left[  m_{0}^{2}-c+\left(  \lambda_{1}+\frac
{\lambda_{2}}{2}\right)  \phi^{2}\right]  \overset{(\ref{minimum})}{=}%
\frac{Z^{2} h_{0}}{\phi}\;,\label{pion}\\
m_{\omega}^{2}  &  =m_{\rho}^{2}=m_{1}^{2}+\frac{\phi^{2}}{2}(h_{1}%
+h_{2}+h_{3})\;,\label{rho}\\
m_{f_{1}}^{2}  &  =m_{a_{1}}^{2}=m_{1}^{2}+g_{1}^{2}\phi^{2}+\frac{\phi^{2}%
}{2} (h_{1}+h_{2}-h_{3})\;\text{.} \label{a1}%
\end{align}
In Appendix A we show the Lagrangian in the form when all shifts have been
explicitly performed. From Eqs.\ (\ref{rho}) and (\ref{a1}) we obtain:
\begin{equation}
m_{a_{1}}^{2}=m_{\rho}^{2}+g_{1}^{2}\phi^{2}-h_{3}\phi^{2}\;\text{.}
\label{massdifference}%
\end{equation}

The pion decay constant, $f_{\pi}$ is determined from the axial current,
\begin{equation}
J_{A_{\mu}}^{a}=\frac{\phi}{Z}\partial_{\mu}\pi^{a}+\ldots\equiv f_{\pi}
\partial_{\mu}\pi^{a}+\ldots\;\; \rightarrow\;\; \phi=Zf_{\pi}\; \text{.}
\label{jA}%
\end{equation}

The large-$N_{c}$ dependence of the parameters is given by
\begin{align}
g_{1},\text{ }g_{2}  &  \propto N_{c}^{-1/2}\;,\nonumber\\
\lambda_{2},\text{ }h_{2},\text{ }h_{3},\text{ }c  &  \propto N_{c}%
^{-1}\;,\nonumber\\
\lambda_{1},\text{ }h_{1}  &  \propto N_{c}^{-2}\;,\nonumber\\
m_{0}^{2},\text{ }m_{1}^{2}  &  \propto N_{c}^{0}\;,\nonumber\\
h_{0}  &  \propto N_{c}^{1/2}\;. \label{largen}%
\end{align}
We remind the reader that a vertex of $n$ quark-antiquark mesons scales as
$N_{c}^{-(n-2)/2}$. As a consequence, the parameters $g_{1}$, $g_{2}$ scale as
$N_{c}^{-1/2}$, because they are associated with a three-point vertex of
quark-antiquark vector fields (of the kind $\rho^{3}$). Similarly, the
parameters $\lambda_{2}$, $h_{2}$, $h_{3}$ scale as $N_{c}^{-1}$, because they
are associated with quartic terms such as $\pi^{4}$ and $\pi^{2}\rho^{2}$. The
parameter $c$ is suppressed by a factor $N_{c}$ although it enters quadratic
mass-like terms. This is due to the fact that the axial anomaly is suppressed
in the large-$N_{c}$ limit. As is evident from Eq.\ (\ref{eta}), the $\eta
_{N}$ meson would also be a Goldstone boson for $N_{c}\rightarrow\infty$. The
parameters $\lambda_{1}$, $h_{1}$ also describe quartic interactions, but are
further suppressed by a factor $1/N_{c}$ because of the trace structure of the
corresponding terms in the Lagrangian. The quantities $m_{0}^{2}$, $m_{1}^{2}$
are mass terms and therefore scale as $N_{c}^{0}$. Then the pion decay
constant $f_{\pi}$ scales as $N_{c}^{1/2}$.\ The quantity $h_{0}$ scales as
$N_{c}^{1/2}$ in order that $m_{\pi}$ scales as $N_{c}^{0}$ as expected. Note
that without any assumptions about the $\sigma$, $a_{0}$, and $f_{1}$, $a_{1}$
fields, we immediately obtain that their masses scale as $N_{c}^{0}$ and their
decay widths as $N_{c}^{-1}$, as we shall see in the following section.
Therefore, they must also correspond to quark-antiquark degrees of freedom.

There are, however, also approaches to the phenomenology of low-lying
axial-vector mesons, such as the one in Ref.\ \cite{Oset1}, where the
Bethe-Salpeter equation is used to unitarize the scattering of vector and
pseudoscalar mesons. Here, the Bethe-Salpeter kernel is given by the lowest-order
effective Lagrangian. This leads to the dynamical generation of resonances,
one of which has a pole mass of 1011 MeV and is consequently assigned to the
$a_{1}(1260)$ meson. This unitarized approach is used in Ref.\ \cite{Oset2} to
study the large-$N_{c}$ behaviour of the dynamically generated resonances,
with the conclusion that the $a_{1}(1260)$ resonance is not a genuine
quark-antiquark state.

However, it was shown in Ref.\ \cite{dynrec} that, while unitarizing the
chiral Lagrangian by means of a Bethe-Salpeter study allows one to find poles
in the complex plane and identify them with physical resonances, it does not
necessarily allow one to make a conclusion about the structure of those
resonances in the large-$N_{c}$ limit. In order to be able to draw correct
conclusions, a Bethe-Salpeter study requires at least one additional term of
higher order not included in the Lagrangian of Refs.\ \cite{Oset1,Oset2}.
Alternatively, the Inverse Amplitude Method of Refs.\ \cite{Pelaez2} can be used.

A very similar approach to the one in Refs.\ \cite{Oset1,Oset2} was also used
in Ref.\ \cite{Leupold} where a very good fit to the $\tau$ decay data from
the ALEPH collaboration \cite{ALEPH} was obtained by fine-tuning the
subtraction point of a loop diagram. Note, however, that detuning the
subtraction point by 5\% will spoil the agreement with experimental data.
Alternately, these data may be described by approaches with the $a_{1}(1260)$
meson as an explicit degree of freedom, such as the one in Ref.\ \cite{UBW},
where $a_{1}(1260)$ is a quark-antiquark state and where the experimental
$a_{1}(1260)$ spectral function is fitted very well. In Ref.\ \cite{UBW},
$m_{a_{1}(1260)}\simeq1150$ MeV and a full width $\Gamma_{a_{1}(1260)}$
$\simeq410$ MeV are obtained. Note that our results, as will be shown later,
give very good results on the $a_{1}(1260)$ phenomenology, for example in the
$a_{1}(1260)\rightarrow\pi\gamma$ and $a_{1}(1260)\rightarrow\rho\pi$ decay
channels, see Sec.\ IV.A.3. 

For the following discussion, it is interesting to note that the $\rho$ meson
mass,
\[
m_{\rho}^{2}=m_{1}^{2}+\frac{\phi^{2}}{2}(h_{1}+h_{2}+h_{3})\;,
\]
can be split into two contributions: the term $m_{1}^{2}$ which does not
depend on the chiral condensate, and the term $\frac{\phi^{2}}{2}(h_{1}
+h_{2}+h_{3})$ which depends quadratically on the condensate and vanishes in
the chirally restored phase. We shall require that none of the two
contributions be negative: in fact, a negative $m_{1}^{2}$ would imply that
the system is unstable when $\phi\rightarrow0$; a negative $\frac{\phi^{2}}%
{2}(h_{1} +h_{2}+h_{3})$ would imply that spontaneous chiral symmetry breaking
decreases the $\rho$ mass. This is clearly unnatural because the breaking of
chiral symmetry generates a sizable effective mass for the light quarks, which
is expected to positively contribute to the meson masses. This positive
contribution is a feature of all known models (such as the Nambu--Jona-Lasinio
model and constituent quark approaches). Indeed, in an important class of
hadronic models (see Ref.\ \cite{harada} and refs.\ therein) the only and
obviously positive contribution to the $\rho$ mass is proportional to
$\phi^{2}$ (i.e., $m_{1}=0$).

\subsection{Equivalent set of parameters}

Instead of the eleven parameters in Eq.\ (\ref{param}), it is technically
simpler to use the following, equivalent set of eleven parameters in the
expressions for the physical quantities:
\begin{equation}
m_{\pi},\text{ }m_{\sigma},\text{ }m_{a_{0}},\text{ }m_{\eta_{N}},\text{ }
m_{\rho},\text{ }m_{a_{1}},\text{ }Z,\text{ }\phi,\text{ }g_{2},\text{ }
h_{1},\text{ }h_{2}\text{.} \label{param2}%
\end{equation}
The quantities $m_{\pi}$, $m_{\rho}$, $m_{a_{1}}$ are taken as the mean values
for the masses of the $\pi$, $\rho$, and $a_{1}$ meson, respectively, as given by
the PDG \cite{PDG}: $m_{\pi}=139.57$ MeV, $m_{\rho}=775.49$ MeV, and
$m_{a_{1}}=1230$ MeV. While $m_{\pi}$ and $m_{\rho}$ are measured to very good
precision, this is not the case for $m_{a_{1}}$. The mass value given above is
referred to as an "educated guess" by the PDG \cite{PDG}. Therefore, we shall also
consider a smaller value, as suggested e.g.\ by the results of
Ref.\ \cite{UBW}. We shall see that, although the overall picture remains
qualitatively unchanged, the description of the decay width of $a_{1}$ into
$\rho\pi$ can be substantially improved.

As outlined in Ref.\ \cite{Mainz},\ the mass of the $\eta_{N}$ meson can be
calculated using the mixing of strange and non-strange contributions in the
physical fields $\eta$ and $\eta^{\prime}(958)$:
\begin{equation}
\eta=\eta_{N}\cos\varphi+\eta_{S}\sin\varphi,\text{ }\eta^{\prime} =-\eta
_{N}\sin\varphi+\eta_{S}\cos\varphi\text{,} \label{phi}%
\end{equation}
where $\eta_{S}$ denotes a pure $\bar{s}s$ state and $\varphi\simeq-36^{\circ
}$ \cite{Giacosa:2007up}. In this way, we obtain the value $m_{\eta_{N}}=716$
MeV. Given the well-known uncertainty of the value of $\varphi$, one could
also consider other values, e.g., $\varphi=-41.4^{\circ}$, as published by the
KLOE Collaboration \cite{KLOE}. In this case, $m_{\eta_{N}}=755$ MeV. The
variation of the $\eta_{N}$ mass does not change the results significantly.

The quantities $\phi$ and $Z$ are linked to the pion decay constant as
$\phi/Z=f_{\pi}=92.4$ MeV. Therefore, the following six quantities remain as
free parameters:
\begin{equation}
m_{\sigma},\text{ }m_{a_{0}},\text{ }Z,\text{ }g_{2},\text{ }h_{1}, \text{
}h_{2}\text{.} \label{param3}%
\end{equation}
The masses $m_{\sigma}$ and $m_{a_{0}}$ depend on the scenario adopted for the
scalar mesons.

At the end of this subsection we report three useful formulas which link the
parameters $g_{1}$, $h_{3}$, and $m_{1}$ of the original set (\ref{param}) to
the second set of parameters (\ref{param2}) [see also Eq.\ (\ref{Z})]:
\begin{align}
g_{1}  &  = g_{1}(Z)=\frac{m_{a_{1}}}{Zf_{\pi}}\sqrt{1-\frac{1}{Z^{2}}%
}\text{,}\label{g1}\\
h_{3}  &  = h_{3}(Z)=\frac{m_{a_{1}}^{2}}{Z^{2}f_{\pi}^{2}}\left(
\frac{m_{\rho}^{2}}{ m_{a_{1}}^{2}} - \frac{1}{Z^{2}}\right)  \text{,}%
\label{h3}\\
m_{1}^{2}  &  = m_{1}^{2}(Z,h_{1},h_{2})=\frac{1}{2}\left[  m_{\rho}%
^{2}+m_{a_{1}}^{2} -Z^{2}f_{\pi}^{2}\left(  g_{1}^{2}+h_{1}+h_{2}\right)
\right]  \text{.} \label{m1eq}%
\end{align}

\section{Decay Widths and $\pi\pi$ Scattering Lengths}

In this section, we quote the formulas for the decay widths and the $\pi\pi$
scattering lengths and specify their dependence on the parameters 
$m_{\sigma}$, $m_{a_{0}}$, $Z$, $g_{2}$, $h_{1}$, and $h_{2}$. 
Using the scaling behavior (\ref{largen}) we obtain that all 
strong decays and scattering lengths scale
as $N_{c}^{-1}$, as expected.

For future use we introduce the momentum function
\begin{equation}
k(m_{a},m_{b},m_{c})=\frac{1}{2 m_{a}}\sqrt{m_{a}^{4}-2 m_{a}^{2} \,
(m_{b}^{2}+m_{c}^{2}) + (m_{b}^{2}-m_{c}^{2})^{2}}\; \theta(m_{a}-m_{b}
-m_{c})\text{.} \label{kabc}%
\end{equation}
In the decay process $a\rightarrow b+c$, with masses $m_{a},\,m_{b},\,m_{c}$,
respectively, the quantity $k(m_{a},m_{b},m_{c})$ represents the modulus of
the three-momentum of the outgoing particles $b$ and $c$ in the rest frame of
the decaying particle $a$. The theta function ensures that the decay width
vanishes below threshold.

\subsection{The \boldmath $\rho\rightarrow\pi\pi$ decay width}

The decay width for $\rho\rightarrow\pi\pi$ reads
\begin{equation}
\Gamma_{\rho\rightarrow\pi\pi}(Z,g_{2})=\frac{m_{\rho}^{5}}{48\pi m_{a_{1}%
}^{4}} \left[  1-\left(  \frac{2m_{\pi}}{m_{\rho}} \right)  ^{2}\right]
^{3/2} \left[  g_{1} Z^{2} + \left(  1-Z^{2}\right)  \frac{g_{2}}{2}\right]
^{2} \text{.} \label{rhopionpion}%
\end{equation}
The experimental value is $\Gamma_{\rho\rightarrow\pi\pi}^{({\rm exp})}=(149.1\pm0.8)$ MeV
\cite{PDG}. The small experimental error can be neglected and the central
value is used as a further constraint allowing us to fix the parameter $g_{2}$
as function of $Z$:
\begin{equation}
g_{2}=g_{2}(Z)=\frac{2}{Z^{2}-1}\left(  g_{1}Z^{2}\pm\frac{4m_{a_{1}}^{2}%
}{m_{\rho}} \sqrt{\frac{3\pi\Gamma_{\rho\rightarrow\pi\pi}^{({\rm exp})}}{(m_{\rho}^{2} -
4m_{\pi}^{2})^{3/2}}} \,\right)  \text{.} \label{g2Z}%
\end{equation}
Note that all input values in Eq.\ (\ref{g2Z}) are experimentally known
\cite{PDG}. The parameter $g_{1}=g_{1}(Z)$ is fixed via Eq.\ (\ref{g1}).

As apparent from Eq.\ (\ref{g2Z}), two solutions for $g_{2}$ are obtained. The
solution with the positive sign in front of the square root may be neglected
because it leads to unphysically large values for the $a_{1}\rightarrow\rho
\pi$ decay width, which is another quantity predicted by our study that also
depends on $g_{2}$ [see Eq.\ (\ref{a1rhopion})]. For example, the value
$Z=1.6$ (see below) would lead to $g_{2}\cong40$ which in turn would give
$\Gamma_{a_{1}\rightarrow\rho\pi}\cong14$ GeV -- clearly an unphysically large
value. Therefore, we will take the solution for $g_{2}$ with the negative sign
in front of the square root. In this case, reasonable values for both $g_{2}$
(see Table~\ref{Table1}) and $\Gamma_{a_{1}\rightarrow\rho\pi}$ (see Sec.\
IV.A.3) are obtained.

\subsection{The \boldmath $f_{1}\rightarrow a_{0}\pi$ decay width}

The decay width $f_{1} \rightarrow a_{0} \pi$ reads
\begin{equation}
\Gamma_{f_{1}\rightarrow a_{0}\pi}(m_{a_{0}},Z,h_{2})=\frac{g_{1}^{2}Z^{2}%
}{2\pi} \frac{k^{3}(m_{f_{1}},m_{a_{0}},m_{\pi})}{m_{f_{1}}^{2}m_{a_{1}}^{4}%
}\left[  m_{\rho}^{2}-\frac{1}{2}(h_{2}+h_{3})\phi^{2}\right]  ^{2}\;\text{.}
\label{f1a0pion}%
\end{equation}

There is a subtle point to comment on here. When the quark-antiquark $a_{0}$
state of our model is identified as the $a_{0}(980)$ meson of the PDG
compilation (Scenario I), then this decay width can be used to fix the
parameter $h_{2}$ as function of $Z,$ $h_{2}\equiv h_{2}(Z)$, by using the
corresponding experimental value $\Gamma_{f_{1}\rightarrow a_{0}\pi}^{({\rm exp})}
=(8.748\pm2.097)$ MeV \cite{PDG}.%
\begin{equation}
h_{2}=h_{2}(Z)=\frac{2}{\phi^{2}}\left(  m_{\rho}^{2}-\frac{h_{3}}{2}\phi
^{2}\pm\frac{m_{f_{1}}m_{a_{0}}^{2}}{g_{1}Z}\sqrt{\frac{2\pi\Gamma
_{f_{1}\rightarrow a_{0}\pi}^{({\rm exp})}}{k^{3}(m_{f_{1}},m_{a_{0}},m_{\pi})}}\,\right)
\text{.} \label{h2Z}%
\end{equation}
Again, there are two solutions, just as in the case of the parameter $g_{2}$.
How strongly the somewhat uncertain experimental value of $\Gamma
_{f_{1}\rightarrow a_{0}\pi}$ influences the possible values of $h_{2}$,
depends on the choice of the sign in front of the square root in
Eq.\ (\ref{h2Z}). Varying $\Gamma_{f_{1}\rightarrow a_{0}\pi}$ within its
experimental range of uncertainty changes the value of $h_{2}$ by an average
of 25\% if the negative sign is chosen, but the same variation of
$\Gamma_{f_{1}\rightarrow a_{0}\pi}$ changes $h_{2}$ by an average of only 6\%
if the positive sign is considered. This is due to the fact that the solution
with the positive square root sign yields larger values of $h_{2}\sim80$,
while the solution with the negative sign leads to $h_{2}\sim20$. The absolute
change of $h_{2}$ is the same in both cases. Our calculations have shown that
using the negative sign in front of the square root yields a too small value
of the $\eta$-$\eta^{\prime}$ mixing angle $\varphi\cong-9^{\circ}$. This
follows by inserting $h_{2}$ into Eq.\ (\ref{a0etapion}) so that it is removed
as a degree of freedom (i.e., replaced by $Z$) and calculating the mixing
angle $\varphi$ from Eq.\ (\ref{a0etapion0}) using the experimental value of the
$a_{0}\rightarrow\eta\pi$ decay amplitude from Ref.\ \cite{Bugg}. For this
reason, we only use the positive sign in front of the square root in
Eq.\ (\ref{h2Z}), i.e., the constraint leading to higher values of $h_{2}$.
Then $\varphi\cong-41.8^{\circ}$ is obtained, in very good agreement with the
central value quoted by the KLOE collaboration \cite{KLOE}, $\varphi
\cong-41.4^{\circ}$ (see also Section IV.A.1).

It may be interesting to note that only the (disregarded) lower value of
$h_{2}$ leads to the expected behaviour of the parameter $h_{1}$ which
[according to Eq.\ (\ref{largen})] should be large-$N_{c}$ suppressed: the
lower value of $h_{2}$ yields $h_{1}=1.8$ whereas the higher value of $h_{2}$
yields $h_{1}=-68$ (see Table\textit{ }\ref{Table1}).

Note that if the quark-antiquark $a_{0}$ meson of our model is identified as
the $a_{0}(1450)$ meson of the PDG compilation (Scenario II) then the
described procedure of replacing $h_{2}$ by $Z$ using Eq.\ (\ref{h2Z}) is no
longer applicable because the decay $f_{1}\rightarrow a_{0}\pi$ is 
kinematically not allowed and its counterpart $a_{0}\rightarrow f_{1}\pi$ has not
been measured.

\subsection{The \boldmath $a_{0}\rightarrow\eta\pi$ and \boldmath $a_{0}%
\rightarrow\eta^{\prime}\pi$ decay amplitudes}

Our $N_{f}=2$ Lagrangian contains the unphysical field $\eta_{N}$. However, by
making use of Eq.\ (\ref{phi}) and invoking the OZI rule, it is possible to
calculate the decay amplitude for the physical process $a_{0}\rightarrow
\eta\pi$ as
\begin{equation}
\label{a0etapion0}A_{a_{0}\eta\pi}=\cos\varphi\;A_{a_{0}\eta_{N}\pi}\text{.}%
\end{equation}
From Eq.\ (\ref{Lagrangian}) the formula for the decay amplitude
containing the non-strange $\eta_{N}$ field is
\begin{equation}
A_{a_{0}\eta_{N}\pi}(m_{a_{0},}Z,h_{2})=
\frac{1}{Zf_\pi} \left\{ m_{\eta_{N}}^{2}-m_{a_{0}}^{2}
+\left( 1 - \frac{1}{Z^2} \right)  \left[  1-\frac{1}{2}
\frac{Z^{2}\phi^{2}}{m_{a_{1}}^{2}}(h_{2}-h_{3})\right]  (m_{a_{0}}%
^{2}-m_{\pi}^{2}-m_{\eta}^{2}) \right\}  \text{.}
\label{a0etapion}%
\end{equation}

Note that Eq.\ (\ref{a0etapion}) contains the unmixed mass $m_{\eta_{N}}$
which enters when expressing the coupling constants in terms of the parameters
(\ref{param2}), as well as the physical mass $m_{\eta}=547.8$ MeV. The latter
arises because the derivative couplings in the Lagrangian lead to the
appearance of scalar invariants formed from the four-momenta of the particles
emerging from the decay, which can be expressed in terms of the physical
(invariant) masses.

The decay width $\Gamma_{a_{0}\rightarrow\eta\pi}$ follows from
Eq.\ (\ref{a0etapion0}) by including a phase space factor:%
\begin{equation}
\Gamma_{a_{0}\rightarrow\eta\pi}(m_{a_{0}},Z,h_{2}) =\frac{k(m_{a_{0}}%
,m_{\eta},m_{\pi})}{8\pi m_{a_{0}}^{2}}\left[  A_{a_{0}\eta\pi}(m_{a_{0}%
},Z,h_{2})\right]  ^{2}\text{.} \label{a0etapion2}%
\end{equation}

In the case of Scenario I, in which $a_{0}\equiv a_{0}(980)$, we shall compare
the decay amplitude $A_{a_{0}\eta\pi}$, Eq.\ (\ref{a0etapion0}), with the
corresponding experimental value deduced from Crystal Barrel data:
$A_{a_{0}\eta\pi}^{({\rm exp})}=(3330\pm150)$ MeV \cite{Bugg}. 
This is preferable to the use
of the decay width quoted by the PDG \cite{PDG} for $a_{0}(980)$, which refers
to the mean peak width, an unreliable quantity due to the closeness of the
kaon-kaon threshold.

In the case of Scenario II, in which $a_{0}\equiv a_{0}(1450)$, it is also
possible to calculate the decay width $a_{0}(1450)\rightarrow\eta^{\prime}\pi
$, using the OZI rule. The amplitude $A_{a_{0}\eta^{\prime}\pi}(m_{a_{0}%
},Z,h_{2})$ is obtained following the same steps as in the previous case,
Eq.\ (\ref{a0etapion}):%
\begin{equation}
A_{a_{0}\eta^{\prime}\pi}(m_{a_{0}},Z,h_{2})= - \frac{\sin\varphi}{Z
f_\pi} \left\{ m_{\eta_{N}}^{2}- m_{a_{0}}^{2}+
\left( 1 - \frac{1}{Z^2} \right) \left[
1-\frac{1}{2}\frac{Z^{2}\phi^{2}}{m_{a_{1}}^{2}}
(h_{2}-h_{3})\right]
(m_{a_{0}}^{2}-m_{\pi}^{2}-m_{\eta^{\prime}}^{2})
\right\}  \;, \label{a0etapion00}
\end{equation}
where the difference compared to Eqs.\ (\ref{a0etapion0}) and 
(\ref{a0etapion}) is the prefactor $- \sin\varphi$ and 
the physical $\eta^{\prime}$ mass
$m_{\eta^{\prime}}=958$ MeV. The corresponding decay width reads:%
\begin{equation}
\Gamma_{a_{0}(1450)\rightarrow\eta^{\prime}\pi}(m_{a_{0}},Z,h_{2})
=\frac{k(m_{a_{0}},m_{\eta^{\prime}},m_{\pi})}{8\pi m_{a_{0}}^{2}} \left[
A_{a_{0}\eta^{\prime}\pi}(m_{a_{0}},Z,h_{2})\right]  ^{2}\text{.}
\label{a0eta'pion2}%
\end{equation}

\subsection{The \boldmath $a_{1}\rightarrow\pi\gamma$ decay width}

We obtain the following formula for the $a_{1}\rightarrow\pi\gamma$ decay
width:%
\begin{equation}
\Gamma_{a_{1}\rightarrow\pi\gamma}(Z)=\frac{e^{2}}{96\pi}\, (Z^{2}-1)\,
m_{a_{1}}\left[  1-\left(  \frac{m_{\pi}}{m_{a_{1}}}\right)  ^{2}\right]
^{3}\text{.} \label{a1piongamma}%
\end{equation}
Note that the $a_{1}\rightarrow\pi\gamma$ decay width depends only on the
renormalisation constant $Z$. Using $\Gamma_{a_{1}\rightarrow\pi\gamma
}^{({\rm exp})}=(0.640\pm0.246)$ MeV \cite{PDG}, one obtains $Z=1.67\pm0.2.$ Unfortunately,
the experimental error for the quantity $\Gamma_{a_{1}\rightarrow\pi\gamma}$
is large. Given that almost all quantities of interest depend very strongly on
$Z$, a better experimental knowledge of this decay would be useful to
constrain $Z$. In the study of Scenario I this decay width will be part of a
$\chi^{2}$ analysis, but still represents the main constraint for $Z$.

\subsection{The \boldmath $\sigma\rightarrow\pi\pi$ decay width}

We obtain the following formula:
\begin{align}
\Gamma_{\sigma\rightarrow\pi\pi}(m_{\sigma},Z,h_{1},h_{2})  &  =\frac{3}{32\pi
m_{\sigma}}\sqrt{ 1 - \left(  \frac{2m_{\pi}}{m_{\sigma}}\right)  ^{2}}
\left\{  \frac{m_{\sigma}^{2} - m_{\pi}^{2}}{Z f_{\pi}} - \frac{g_{1}^{2}%
Z^{3}f_{\pi}}{m_{a_{1}}^{4}}\left[  m_{\rho}^{2} - \frac{\phi^{2}}{2}%
(h_{1}+h_{2}+h_{3})\right]  (m_{\sigma}^{2}- 2\, m_{\pi}^{2})\right\}  ^{2}\;.
\label{sigmapionpion}%
\end{align}
It is apparent from Eqs.\ (\ref{largen}) that the sigma decay width decreases
as the number of colours $N_{c}$ increases. Thus, the sigma field in our model
is a $\bar{q}q$ state \cite{Pelaez2}. In Scenario I we have assigned the
$\sigma$ field as $f_{0}(600)$, correspondingly we are working with the
assumption that $f_{0}(600)$ [as well as $a_{0}(980)$] is a $\bar{q}q$ state.
In Scenario II, the same assumption is valid for the $f_{0}(1370)$ and
$a_{0}(1450)$ states.

Note that in Eq.\ (\ref{sigmapionpion}) the first term in braces arises from
the scalar $\sigma\pi\pi$ vertex, while the second term comes from the
coupling of the $\sigma$ to the $a_{1}$, which becomes a derivatively coupled
pion after the shift (\ref{shifts}). Because of the different signs, these two
terms interfere destructively. As the decay width of a light $\sigma$ meson
into two pions can be very well reproduced in the linear sigma model without
vector mesons (corresponding to the case $g_{1} \rightarrow0$), this
interference prevents obtaining a reasonable value for this decay width in the
present model with vector mesons, see Sec.\ IV.A.2. This problem does not
occur for a heavy $\sigma$ meson, see Sec.\ IV.B.3 and Ref.\ \cite{Zakopane}.

\subsection{The \boldmath $a_{1}\rightarrow\sigma\pi$ decay width}

The formula for the decay width reads%
\begin{equation}
\Gamma_{a_{1}\rightarrow\sigma\pi}(m_{\sigma},Z,h_{1},h_{2})=\frac
{k^{3}(m_{a_{1}},m_{\sigma},m_{\pi})}{6\pi m_{a_{1}}^{6}}g_{1}^{2}Z^{2}\left[
m_{\rho}^{2}-\frac{\phi^{2}}{2}(h_{1}+h_{2}+h_{3})\right]  ^{2}\text{.}
\label{a1sigmapion}%
\end{equation}

\subsection{The \boldmath $a_{1}\rightarrow\rho\pi$ decay width}

Let $P$ be the four-momentum of the $a_{1}$ meson, $K_{1}$ the four-momentum
of the $\rho$ meson and $K_{2}$ the four-momentum of the pion. Then the
following formula for the $a_{1}\rightarrow\rho\pi$ decay width is obtained:
\begin{equation}
\Gamma_{a_{1}\rightarrow\rho\pi}(Z)=\frac{k(m_{a_{1}},m_{\rho},m_{\pi})}{12\pi
m_{a_{1}}^{2}}\left[  (h_{\mu\nu})^{2}-\frac{(h_{\mu\nu}K_{1}^{\nu})^{2}%
}{m_{\rho}^{2}} -\frac{(h_{\mu\nu}P^{\mu})^{2}}{m_{a_{1}}^{2}}+\frac
{(h_{\mu\nu}P^{\mu}K_{1}^{\nu})^{2}}{m_{\rho}^{2}m_{a_{1}}^{2}}\right]  \text{
,} \label{a1rhopion}%
\end{equation}

where $h_{\mu\nu}$ is the vertex following from the relevant part of the
Lagrangian (\ref{Lagrangian}) that reads%
\[
h_{\mu\nu}=Z^{2} f_{\pi}\left\{  (g_{1}^{2}-h_{3})\,g_{\mu\nu} +\frac
{g_{1}g_{2}}{m_{a_{1}}^{2}}[K_{1\mu}K_{2\nu}+K_{2\mu}P_{\nu}-K_{2}\cdot
(K_{1}+P)g_{\mu\nu}] \right\}
\]

and
\begin{align*}
K_{1}\cdot K_{2}  &  =\frac{m_{a_{1}}^{2}-m_{\rho}^{2}-m_{\pi}^{2}}{2}\;,\\
P\cdot K_{2}  &  =m_{a_{1}}E_{\pi}\equiv m_{a_{1}}\sqrt{k^{2}+m_{\pi}^{2}}\;.
\end{align*}

Thus, we have
\begin{align*}
h_{\mu\nu}^{2}  &  =Z^{4}f_{\pi}^{2}\left\{  4(g_{1}^{2}-h_{3})^{2}%
+\frac{g_{1}^{2}g_{2}^{2}}{m_{a_{1}}^{4}}\left[  m_{a_{1}}^{4}+m_{\pi}%
^{4}+m_{\rho}^{4}+m_{\pi}^{2}m_{\rho}^{2}+m_{a_{1}}^{2}(m_{\pi}^{2}-2m_{\rho
}^{2})+3(m_{a_{1}}^{2}-m_{\rho}^{2}-m_{\pi}^{2})m_{a_{1}}E_{\pi}\right]
\right. \\
&  \hspace*{1.3cm}-\left.  3\,\frac{g_{1}g_{2}(g_{1}^{2}-h_{3})}{m_{a_{1}}%
^{2}}\left(  m_{a_{1}}^{2}-m_{\rho}^{2}-m_{\pi}^{2}+2m_{a_{1}}E_{\pi}\right)
\right\}  \text{,}\\
(h_{\mu\nu}K_{1}^{\nu})^{2}  &  =Z^{4}f_{\pi}^{2}\left\{  (g_{1}^{2}%
-h_{3})^{2}m_{\rho}^{2}+\frac{g_{1}^{2}g_{2}^{2}}{4m_{a_{1}}^{4}}[(m_{\pi}%
^{2}-m_{\rho}^{2})^{2}(m_{\pi}^{2}+m_{\rho}^{2}-2m_{a_{1}}^{2})+(m_{\pi}%
^{2}+m_{\rho}^{2})m_{a_{1}}^{4}\right. \\
&  \hspace*{1.3cm}\left.  -4(m_{a_{1}}^{2}-m_{\rho}^{2}-m_{\pi}^{2})m_{a_{1}%
}^{2}E_{\rho}E_{\pi}]+\frac{g_{1}g_{2}(g_{1}^{2}-h_{3})}{m_{a_{1}}^{2}}\left[
(m_{a_{1}}^{2}-m_{\pi}^{2})m_{a_{1}}E_{\rho}-2m_{a_{1}}m_{\rho}^{2}\left(
E_{\pi}+\frac{E_{\rho}}{2}\right)  \right]  \right\}  \text{,}\\
(h_{\mu\nu}P^{\mu})^{2}  &  =Z^{4}f_{\pi}^{2}\left\{  (g_{1}^{2}-h_{3}%
)^{2}m_{a_{1}}^{2}+\frac{g_{1}^{2}g_{2}^{2}}{4m_{a_{1}}^{4}}[(m_{a_{1}}%
^{2}-m_{\pi}^{2})^{2}(m_{a_{1}}^{2}+m_{\pi}^{2}-2m_{\rho}^{2})+(m_{\pi}%
^{2}+m_{a_{1}}^{2})m_{\rho}^{4}\right. \\
&  \hspace*{1.3cm}\left.  -4(m_{a_{1}}^{2}-m_{\rho}^{2}-m_{\pi}^{2})m_{a_{1}%
}^{2}E_{\rho}E_{\pi}]+\frac{g_{1}g_{2}(g_{1}^{2}-h_{3})}{m_{a_{1}}^{2}%
}[2m_{a_{1}}^{2}E_{\rho}E_{\pi}-(m_{a_{1}}^{2}-m_{\rho}^{2}-m_{\pi}%
^{2})m_{a_{1}}^{2}]\right\}  \;,\\
(h_{\mu\nu}P^{\mu}K_{1}^{\nu})^{2}  &  =(g_{1}^{2}-h_{3})^{2}Z^{4}f_{\pi}%
^{2}m_{a_{1}}^{2}E_{\rho}^{2}\;\text{.}%
\end{align*}

\subsection{The tree-level scattering lengths}

The partial wave decomposition \cite{ACGL} leads to the following formula for
the s-wave $I=0$ pion-pion scattering length $a_{0}^{0}$ (in units of $m_{\pi
}^{-1}$):%
\begin{align}
a_{0}^{0}(Z,m_{\sigma},h_{1})  &  =\frac{1}{4\pi}\left(  2g_{1}^{2}Z^{4}%
\frac{m_{\pi}^{2}}{m_{a_{1}}^{4}}\left\{  m_{\rho}^{2}+\frac{\phi^{2}}%
{16}[12g_{1}^{2}-2(h_{1}+h_{2})-14h_{3}]\right\}  -\frac{5}{8}\frac
{Z^{2}m_{\sigma}^{2}-m_{\pi}^{2}}{f_{\pi}^{2}}\right. \nonumber\\
&  \hspace*{1cm} -\frac{3}{2}\left\{  g_{1}^{2}Z^{2}\phi\frac{m_{\pi}^{2}%
}{m_{a_{1}}^{4}}\left[  2m_{a_{1}}^{2} +m_{\rho}^{2}-\frac{\phi^{2}}{2}
(h_{1}+h_{2}+h_{3})\right]  -\frac{Z^{2}m_{\sigma}^{2}-m_{\pi}^{2}}{2\phi
}\right\}  ^{2}\frac{1}{4m_{\pi}^{2}-m_{\sigma}^{2}}\nonumber\\
&  \hspace*{1cm} +\left.  \left\{  g_{1}^{2}Z^{2}\phi\frac{m_{\pi}^{2}%
}{m_{a_{1}}^{4}}\left[  m_{\rho}^{2}-\frac{\phi^{2}}{2}(h_{1}+h_{2}%
+h_{3})\right]  +\frac{Z^{2}m_{\sigma}^{2}-m_{\pi}^{2}}{2\phi}\right\}
^{2}\frac{1}{m_{\sigma}^{2}}\right)  \text{.} \label{a00}%
\end{align}
We use the value $a_{0}^{0\,({\rm exp})}=0.218\pm0.020$ in accordance with the 2003 and
2004 data from the NA48/2 collaboration \cite{Peyaud}.

An analogous calculation leads to the s-wave $I=2$ pion-pion scattering length
$a_{0}^{2}$:
\begin{align}
a_{0}^{2}(Z,m_{\sigma},h_{1})  &  =-\frac{1}{4\pi}\left(  \frac{Z^{2}%
m_{\sigma}^{2}-m_{\pi}^{2}}{4 f_{\pi}^{2}}+g_{1}^{2}Z^{4}\frac{m_{\pi}^{2}%
}{m_{a_{1}}^{4}}\left[  m_{\rho}^{2}-\frac{\phi^{2}}{2}(h_{1}+h_{2}%
+h_{3})\right]  \right. \nonumber\\
&  \hspace*{1cm} -\left.  \left\{  g_{1}^{2}Z^{2}\phi\frac{m_{\pi}^{2}%
}{m_{a_{1}}^{4}}\left[  m_{\rho}^{2}-\frac{\phi^{2}}{2}(h_{1}+h_{2}%
+h_{3})\right]  +\frac{Z^{2}m_{\sigma}^{2}-m_{\pi}^{2}}{2\phi}\right\}
^{2}\frac{1}{m_{\sigma}^{2}}\right)  \text{.} \label{a02}%
\end{align}
The experimental result for $a_{0}^{2}$ from the NA48/2 collaboration is
$a_{0}^{2\,({\rm exp})}=-0.0457\pm0.0125$ \cite{Peyaud}.
Note that the $\pi\pi$
scattering lengths were also studied away from threshold 
in Ref.\ \cite{Schechter}, in a model quite similar to ours.

\section{Study of Different Scenarios for the Structure of Scalar Mesons}

In this section we discuss two different interpretations of the scalar mesons.
The following subsection describes the results obtained when $f_{0}(600)$ and
$a_{0}(980)$ are interpreted as scalar quarkonia (Scenario I). Then we discuss
the results obtained when $f_{0}(1370)$ and
$a_{0}(1450)$ are interpreted as scalar quarkonia (Scenario II).

\subsection{Scenario I: Light Scalar Quarkonia}

\subsubsection{Fit procedure}

As a first step we utilize the central value of the experimental result
$\Gamma_{\rho\rightarrow\pi\pi}^{({\rm exp})}=149.1$ MeV \cite{PDG} in order to
express the parameter $g_{2}$ as a function of $Z$ via Eq.\ (\ref{g2Z}). 
Moreover, we fix the mass
$m_{a_{0}}=0.98$ GeV \cite{PDG} and we also use the central value
$\Gamma_{f_{1}\rightarrow a_{0}\pi}(Z,h_{2})=8.748$ MeV to express $h_{2}$ as
a function of $Z$. The results are practically unaffected by the 6\%
uncertainty in $h_2$ originating from the uncertainty in 
$\Gamma_{f_{1}\rightarrow a_{0}\pi}$, see Eq.\ (\ref{h2Z}). 

As a result, the set of free parameters in Eq.\ (\ref{param3}) is further
reduced to three parameters:
\begin{equation}
Z,\text{ }m_{\sigma},\text{ }h_{1}\;.
\end{equation}
Note that in this scenario the field $\sigma$ is identified with the resonance
$f_{0}(600)$, but the experimental uncertainty on its mass is so large that it
does not allow us to fix $m_{\sigma}$. We therefore keep $m_{\sigma}$ as a
free parameter.

We now determine the parameters $Z$, $h_{1}$, and $m_{\sigma}$ using known
data on the $a_{1}\rightarrow\pi\gamma$ decay width (\ref{a1piongamma}) and on
the $\pi\pi$ scattering lengths $a_{0}^{0}$ and $a_{0}^{2}$ reported in
Eqs.\ (\ref{a00}) and (\ref{a02}). This is a system of three equations with
three variables and can be solved uniquely. We make use of the $\chi^{2}$
method in order to determine not only the central values for our parameters
but also their error intervals:%

\begin{equation}
\chi^{2}(Z,m_{\sigma},h_{1})=\left(  \frac{\Gamma_{a_{1}\rightarrow\pi\gamma
}(Z)-\Gamma_{a_{1}\rightarrow\pi\gamma}^{\mathrm{(exp)}}}{\triangle
\Gamma_{\mathrm{decay}}^{\mathrm{(exp)}}}\right)  ^{2}+\sum_{i\in
\{0,2\}}\left(  \frac{a_{0}^{i}(Z,m_{\sigma},h_{1})-a_{0}^{i\mathrm{\,(exp)}%
}}{\triangle a_{0}^{i\mathrm{\,(exp)}}}\right)  ^{2}. \label{chi}%
\end{equation}
The errors for the model parameters are calculated as the square roots of the
diagonal elements of the inverted Hessian matrix obtained from $\chi
^{2}(Z,m_{\sigma},h_{1})$. The minimal value is obtained for
$\chi^{2}=0$, as expected given that the parameters are determined from a
uniquely solvable system of equations. The values of the parameters are as
follows:
\begin{equation}
Z=1.67\pm0.2\,,\;m_{\sigma}=(332\pm456)\text{ MeV\,,\; }h_{1}=-68\pm338\;.
\label{ksquared}%
\end{equation}

\begin{figure}[h]
\begin{center}
\includegraphics[width=10cm]{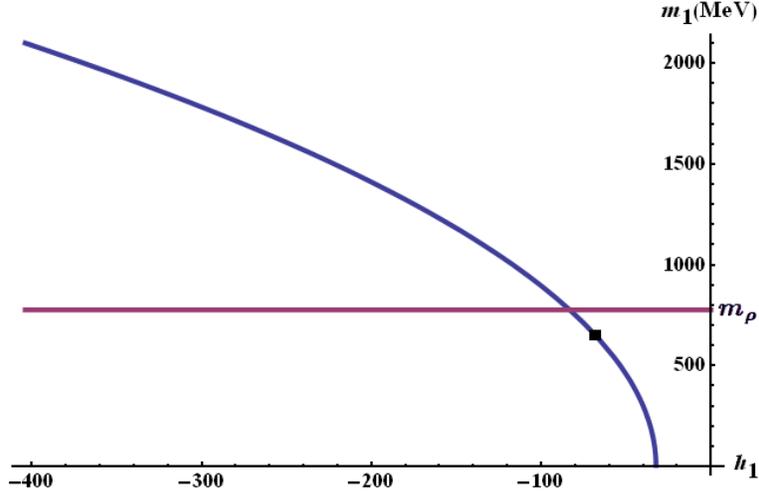}
\end{center}
\caption{$m_{1}$ as function of $h_{1}$, constrained at the central value of
$Z=1.67$. The black dot marks the position of central values $h_{1}=-68$ and
$m_{1}=652$ MeV.}%
\label{m1f}%
\end{figure}

Clearly, the error intervals for $m_{\sigma}$ and $h_{1}$ are very large.
Fortunately, it is possible to constrain the $h_{1}$ error interval as
follows. As described at the end of Sec.\ II.B,
the $\rho$ mass squared contains two contributions -- \ the bare mass term
$m_{1}^{2}$ and the quark condensate contribution ($\sim\phi^{2}$). The
contribution of the quark condensate is special for the globally invariant
sigma model; in the locally invariant model $m_{\rho}$ is always equal to
$m_{1}$ \cite{RS}. Each of these contributions should have at most the value
of 775.49 MeV ($=m_{\rho}$) because otherwise either the bare mass or the
quark condensate contribution to the rho mass would be negative, which appears
to be unphysical. A plot of the function 
$m_{1}=m_{1}(Z,h_{1},h_{2}(Z)),$ see Eq.\ (\ref{m1eq}),
for the central values of $Z=1.67$ and $\Gamma_{f_{1}\rightarrow a_{0}\pi
}^{({\rm exp})}=8.748$ MeV is shown in Fig.\ \ref{m1f}. 

Note that varying the value of $\Gamma_{f_{1}\rightarrow
a_{0}\pi}^{({\rm exp})}$ within its
experimental boundaries would only very slightly change $h_{1}$ by $\pm4$ and
this parameter is thus unaffected by the experimental error for $\Gamma
_{f_{1}\rightarrow a_{0}\pi}^{({\rm exp})}$. If the value of $m_{1}$ was known exactly, then
Eq.\ (\ref{m1eq}) would allow us to constrain $h_{1}$ via $Z$. However, given
that at this point we can only state that $0\leq m_{1}\leq m_{\rho}$, for each
$Z$ one may consider all values of $h_{1}$ between two boundaries, one
obtained from the condition $m_{1}(Z,h_{1},h_{2}(Z))\equiv0$ and another
obtained from the condition $m_{1}(Z,h_{1},h_{2}(Z))\equiv m_{\rho}$. For
example, using the central value of $Z=1.67$, we obtain $-83\leq h_{1}\leq
-32$. The lower boundary follows from $m_{1}\equiv m_{\rho}$ and the upper
boundary from $m_{1}\equiv0$, see Fig.\ \ref{m1f}. Note that the central value
$h_{1}=-68$ from Eq.\ (\ref{ksquared}) corresponds to $m_{1}=652$ MeV. If the
minimal value of $Z=1.47$ is used, then $h_{1}=-112$ is obtained from
$m_{1}\equiv m_{\rho}$ and $h_{1}=-46$ from $m_{1}\equiv0$. Thus, $-112\leq
h_{1}\leq-46$ for $Z=1.47$. Analogously, $-64\leq h_{1}\leq-24$ is obtained
for the maximal value $Z=1.87$.

Clearly, each lower boundary for $h_{1}$ is equivalent to $m_{1}\equiv
m_{\rho}$ and each upper boundary for $h_{1}$ is equivalent to $m_{1}\equiv0$.
Thus, in the following we will only state the values of $Z$ and $m_{1}$;
$h_{1}$ can always be calculated using Eq.\ (\ref{m1eq}). In this way, the
dependence of our results on $m_{1}$ and thus on the origin of the $\rho$ mass
will be exhibited.

\begin{figure}
[h]
\begin{center}
\includegraphics[
height=2.162in,
width=6.8934in
]%
{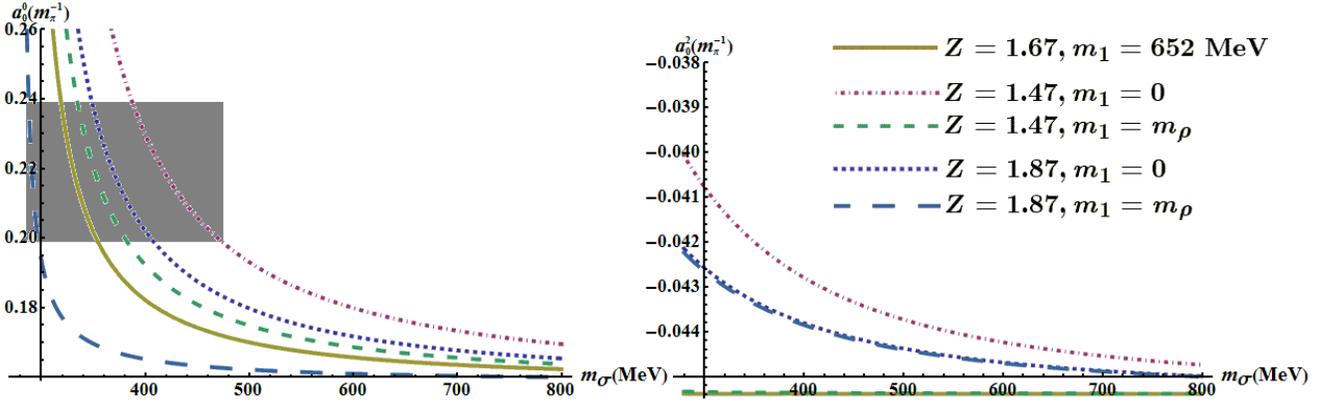}%
\caption{Scattering lengths $a_{0}^{0}$ and $a_{0}^{2}$ as function of
$m_{\sigma}$ [the shaded band corresponds to the NA48/2 value of $a_{0}^{0}$;
no error interval is shown for $a_{0}^{2}$ due to the large interval size
\cite{Peyaud}].}%
\label{a00a02f}%
\end{center}
\end{figure}

The value of $m_{\sigma}$ can be constrained in a way similar to $h_{1}$ using
the scattering length $a_{0}^{0}$; the scattering length $a_{0}^{2}$ possesses
a rather large error interval making it unsuitable to constrain $m_{\sigma}$.
Figure \ref{a00a02f} shows the different values for $a_{0}^{0}$ and $a_{0}%
^{2}$ depending on the choice of $Z$ and $m_{1}$.%

It is obvious that the value of $a_{0}^{0}$ is only consistent with the NA48/2
value \cite{Peyaud}, if $m_{\sigma}$ is in the interval [288, 477] MeV, i.e.,
$m_{\sigma}=332_{-44}^{+145}$ MeV. This value for $m_{\sigma}$ follows if the
parameters $Z$ and $m_{1}$ are varied within the allowed boundaries. If we
only consider the $a_{0}^{0}$ curve that is obtained for the central values of
$Z$ and $m_{1}$, a much more constrained value of $m_{\sigma}=332_{-13}^{+24}$
MeV follows from Fig.\ \ref{a00a02f}. We will be working with the broader
interval of $m_{\sigma}$. Even then, constraining $m_{1}$ to the interval
$[0,m_{\rho}]$, the error bars for $m_{\sigma}$ are reduced by at least a
factor of three in comparison to the result (\ref{ksquared}) following from
the $\chi^{2}$ calculation.

We summarize our results for the parameters $Z$ and $m_{\sigma}$:
\[
Z=1.67\pm0.2\,, \;m_{\sigma}=332_{-44}^{+145}\text{ MeV\;.}%
\]
The central values of all parameters of the original set (\ref{param}) are
given in Table\textit{ }\ref{Table1}. They follow from the $\chi^{2}$\ fit
($m_{\sigma}$, $h_{1}$), via decay width constraints ($h_{2}$, $g_{2}$), and
from Eqs.\ (\ref{sigma}) -- (\ref{a1}) and (\ref{g1}) -- (\ref{h3}). The
central values of $Z$, $m_{\sigma}$, and $h_{1}$, Eq.\ (\ref{ksquared}), have
been used to calculate all other parameters. We neglect the errors, apart from
those of $m_{1}$, which in this scenario vary in a large range.

\bigskip%
\begin{table}[h] \centering
\begin{tabular}
[c]{|c|c|c|c|c|c|c|c|c|c|c|c|c|}\hline
\textit{Parameter} & $m_{\sigma}$ & $h_{1}$ & $h_{2}$ & $h_{3}$ & $g_{1}$ &
$g_{2}$ & $m_{0}$ & $m_{1}$ & $\lambda_{1}$ & $\lambda_{2}$ & $c$ & $h_{0}%
$\\\hline
\textit{Value} & 332 MeV & -68 & 80 & 2.4 & 6.4 & 3.1 & 210 MeV &
652$_{-652}^{+123}$ MeV & -14 & 33 & 88744 MeV$^{2}$ & $1\cdot10^{6}$
MeV$^{3}$\\\hline
\end{tabular}
\caption{Central values of parameters for Scenario I.}\label{Table1}%
\end{table}%

Note that the values of $a_{0}^{2}$ depend strongly on the choice of the
parameters $Z$ and $m_{1}$. Whereas for the central values of $Z$ and $m_{1}$
this scattering length is constant and has the value $a_{0}^{2}=-0.0454$, its
value increases if $Z$ and $m_{1}$ are considered at their respective
boundaries, see Fig.\ \ref{a00a02f}.

The value of $Z$ alone allows us to calculate certain decay widths in the
model. For example, as a consistency check we obtain $\Gamma_{a_{1}%
\rightarrow\pi\gamma}=0.640_{-0.231}^{+0.261}$ MeV which is in good agreement
with the experimental result. Also, given that the $a_{0}\rightarrow\eta
_{N}\pi$ decay amplitude only depends on $Z$, it is possible to calculate the
value of this amplitude, Eq. (\ref{a0etapion}). For $Z=1.67$, we obtain
$A_{a_{0}\rightarrow\eta\pi}=3939$ MeV for the decay amplitude $a_{0}%
\rightarrow\eta\pi$ involving the physical $\eta$ field if the $\eta$%
-$\eta^{\prime}$ mixing angle of $\varphi=-36%
{{}^\circ}%
$ \cite{Giacosa:2007up} is taken. The Crystal Barrel data \cite{Bugg} read
$A_{a_{0}\rightarrow\eta\pi}^{({\rm exp})}=3330$ MeV and hence there is an approximate
discrepancy of 20\%. If the KLOE Collaboration \cite{KLOE} value of
$\varphi=-41.4^{\circ}$ is considered, then the value of 
$A_{a_{0}\rightarrow\eta\pi}=3373$ MeV
follows -- in perfect agreement with the Crystal Barrel value. From this we
conclude that this scenario prefers a relatively large value of the $\eta
$-$\eta^{\prime}$ mixing angle. In fact, if we use the Crystal Barrel
value $A_{a_{0}\rightarrow\eta\pi}^{({\rm exp})}=3330$ MeV as input,
we would predict $\varphi=-41.8^\circ$ for the central value of $Z$ 
as well as $\varphi=-42.3^\circ$ and $\varphi=-41.6^\circ$
for the highest and lowest values of $Z$, respectively, i.e., $\varphi
=-41.8^\circ{}_{-0.5^\circ}^{+0.2^\circ}$. This is in excellent agreement with the KLOE
collaboration result $\varphi=-41.4^{\circ} \pm 0.5^{\circ}$.

\subsubsection{The decay $\sigma\rightarrow\pi\pi$}

\begin{figure}
[h]
\begin{center}
\includegraphics[
height=2.1326in,
width=4.8369in
]%
{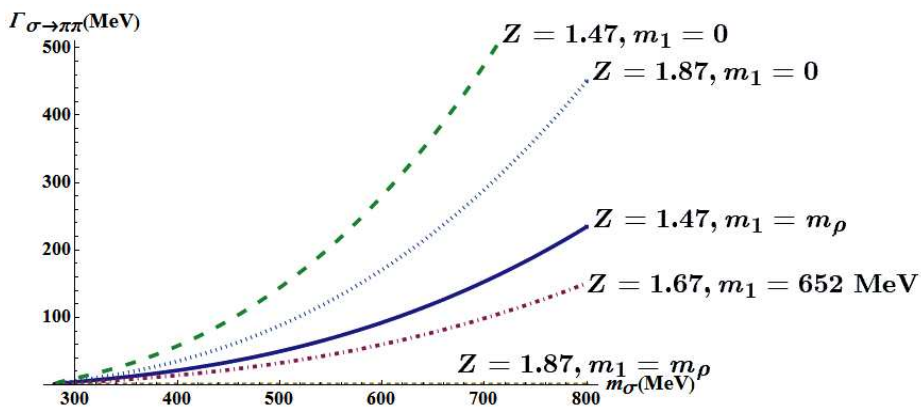}%
\caption{{}$\Gamma_{\sigma\rightarrow\pi\pi}$ as function of $m_{\sigma}$ for
different values of $Z$ and $m_{1}$. The PDG \cite{PDG} quotes $\Gamma_{\sigma
\rightarrow \pi\pi}^{({\rm exp})}=(600-1000)$ MeV; 
the results from the chiral perturbation theory suggest
$\Gamma_{\sigma \rightarrow \pi\pi}=544$ MeV \cite{Leutwyler} and 
$\Gamma_{\sigma \rightarrow \pi\pi}=510$ MeV \cite{Pelaez1}.}%
\label{Sigmaf1}%
\end{center}
\end{figure}

The sigma decay width $\Gamma_{\sigma\rightarrow\pi\pi}$ depends on all three
parameters $Z$, $m_{1}$ (originally $h_{1}$), and $m_{\sigma}$. In
Fig.\ \ref{Sigmaf1} we show the dependence of this decay width on the sigma
mass for fixed values of $Z$ and $m_{1}$, varying the latter within their
respective boundaries.

\bigskip%

Generally, the values that we obtain are too small when compared to the PDG
data \cite{PDG} and to other calculations of the sigma meson decay width, such
as the one performed by Leutwyler \textit{et al.} \cite{Leutwyler} who found
$\Gamma_{\sigma \rightarrow \pi\pi}/2=272_{-12.5}^{+9}$ MeV and Pel\'{a}ez \textit{et al.}
\cite{Pelaez1}\ who found $\Gamma_{\sigma \rightarrow \pi
\pi}/2=(255\pm16)$ MeV. The largest
values for the decay width that we were able to obtain within our model are
for the case when $Z$ is as small as possible, $Z=1.47$, and $m_{1}=0$, i.e.,
when the $\rho$ mass is solely generated by the quark condensate. As seen
above, for this case the scattering lengths allow a maximum value $m_{\sigma}=
477$ MeV, for which $\Gamma_{\sigma\rightarrow\pi\pi}\cong145$ MeV. In all
other cases, the decay width is even smaller. However, as will be discussed in
Sec.\ IV.A.3, the case $m_{1}=0$ leads to the unphysically small value
$\Gamma_{a_{1}\rightarrow\sigma\pi} \simeq0$ and should therefore not be taken
too seriously. As apparent from Fig.\ \ref{a00a02f}, excluding small values of
$m_{1}$ would require smaller values for $m_{\sigma}$ in order to be
consistent with the scattering lengths. According to Fig.\ \ref{Sigmaf1},
however, this in turn leads to even smaller values for the decay width.

Hence, we conclude that the isoscalar meson in our model cannot be
$f_{0}(600)$, thus excluding that this resonance is predominantly a $\bar{q}q$ state and the
chiral partner of the pion. Then the interpretation of the isospin-one state
$a_{0}(980)$ as a (predominantly) quarkonium state is also excluded. The only choice
is to consider Scenario II, see Sec.\ IV.B, i.e., to interpret the scalar states
above 1 GeV, $f_{0}(1370)$ and $a_{0}(1450)$, as being predominantly
quarkonia. If the decay width
of $f_{0}(1370)$ can be described by the model, this would be a very strong
indication that these higher-lying states can be indeed interpreted as
(predominantly) $\bar{q}q$ states. Note that very similar results about the
nature of the light scalar mesons were also found using different approaches: from
an analysis of the meson behaviour in the large-$N_{c}$ limit in
Refs.\ \cite{Pelaez2} and \cite{Sannino} as well as from lattice studies, such
as those in Refs.\ \cite{Liu}.

We remark that the cause for preventing a reasonable fit of the light sigma
decay width is the interference term arising from the vector mesons in
Eq.\ (\ref{sigmapionpion}). In the unphysical case without vector
meson degrees of freedom, a simultaneous fit of the
decay width and the scattering lengths is possible
\cite{Zakopane}.

\subsubsection{Decays of the $a_{1}$ meson}

\begin{figure}
[h]
\begin{center}
\includegraphics[
height=2.6299in,
width=4.4884in
]%
{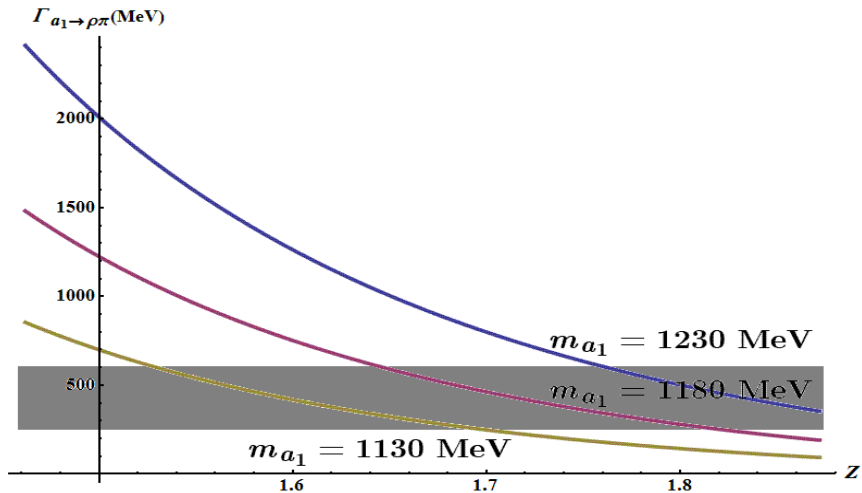}%
\caption{$\Gamma_{a_{1}\rightarrow\rho\pi}$ for different values of $m_{a_{1}%
}$. The shaded area corresponds to the possible values of $\Gamma
_{a_{1}\rightarrow\rho\pi}$ as stated by the PDG. }%
\label{a1rhopif}%
\end{center}
\end{figure}

We first consider the decay width $\Gamma_{a_{1} \rightarrow\rho\pi}$. For a
given $m_{a_{1}}$, this decay width depends only on $Z$. The PDG quotes a
rather large band of values, $\Gamma_{a_{1}\rightarrow\rho\pi}^{\mathrm{(exp)}%
}=(250-600)$ MeV. For $m_{a_{1}}=1230$ MeV, our fit of meson properties yields
$Z=1.67 \pm0.2$. The ensuing region is shown as shaded area in
Fig.\ \ref{a1rhopif}. For $m_{a_{1}}=1230$ MeV, $\Gamma_{a_{1}\rightarrow
\rho\pi}$ decreases from 2.4 GeV to 353 MeV, if $Z$ varies from 1.47 to 1.87.

We also observe from Fig.\ \ref{a1rhopif} that the range of values for $Z$,
which give values for $\Gamma_{a_{1} \rightarrow\rho\pi}$ consistent with the
experimental error band, becomes larger if one considers smaller masses for
the $a_{1}$ meson. We have taken $m_{a_{1}}=1180$ MeV and $m_{a_{1}}=1130$
MeV, the latter being similar to the values used in Refs.\ \cite{UBW} and
\cite{Williams:2009rk}. Repeating our calculations, we obtain a new range of
possible values for $Z$, $Z \simeq1.69 \pm0.2$ for $m_{a_{1}} = 1180$ MeV and
$Z \simeq1.71. \pm0.2$ for $m_{a_{1}} = 1130$ MeV. For the respective central
values of $Z$ we then compute $\Gamma_{a_{1}\rightarrow\rho\pi}^{m_{a_{1}%
}=1180\mathrm{\ MeV}}=483$ MeV ($Z^{m_{a_{1} }=1180\,\mathrm{MeV}}=1.69$) and
$\Gamma_{a_{1}\rightarrow\rho\pi}^{m_{a_{1}}=1130\mathrm{\ MeV}}=226$ MeV
($Z^{m_{a_{1}}=1130\,\mathrm{MeV}}=1.71$), in good agreement with experimental
data. All other results remain valid when $m_{a_{1}}$ is decreased by about
100 MeV. Most notably, the $f_{0}(600)$ decay width remains too small.

\bigskip%

We also consider the $a_{1}\rightarrow\sigma\pi$ decay width. Experimental
data on this decay channel \cite{PDG} are inconclusive. The value
$\Gamma_{a_{1}\rightarrow\sigma\pi}=56$ MeV is obtained for the central values
of $Z$, $m_{1}$, $m_{\sigma}$, and $\Gamma_{f_{1}\rightarrow a_{0}\pi}$ (which
was used to constrain $h_{2}$ via $Z$). Taking the limit $m_{1}=0$ pulls the
value of $\Gamma_{a_{1}\rightarrow\sigma\pi}$ down to practically zero,
regardless whether $Z=Z_{\min}$ or $Z=Z_{\max}$. This is an indication that
the $m_{1}=0$ limit, where $m_{\rho}$ is completely generated from the quark
condensate, cannot be physical. Note that the case $Z=Z_{\max}=1.87$ and
$m_{1}\equiv m_{\rho}$, i.e., where the quark condensate contribution to the
$\rho$ mass vanishes, leads to a rather large value of $\Gamma_{a_{1}%
\rightarrow\sigma\pi}$, e.g., for the central value of $m_{\sigma}=332$ MeV
the value of $\Gamma_{a_{1}\rightarrow\sigma\pi}=120$ MeV follows.
Interestingly, this picture persists even if lower values of $m_{a_{1}}$ are
considered. Improving experimental data for this decay channel would allow us
to further constrain our parameters.

\subsubsection{\textit{The case of isospin-exact scattering lengths.}}

So far, the values of the scattering lengths used in our fit, $a_{0}^{0}%
=0.218\pm0.020$ and $a_{0}^{2}=-0.0457\pm0.0125$ \cite{Peyaud}, account for
the small explicit breaking of isospin symmetry due to the difference of the
up and down quark masses. However, in our model the isospin symmetry is exact.
Thus, one should rather use the isospin-exact values
$a_{0}^{0\,\mathrm{(I)}}=0.244\pm0.020$ and 
$a_{0}^{2\,\mathrm{(I)}}=-0.0385\pm0.0125$ \cite{Bloch}.
In this section we will briefly show that the conclusions reached so
far remain qualitatively unchanged if the isospin-exact values for the scattering 
lengths are considered.

Performing the $\chi^{2}$ fit, Eq.\ (\ref{chi}), with $\Gamma_{a_{1}
\rightarrow\pi\gamma}$, $a_{0}^{0\,\mathrm{(I)}}$ and
$a_{0}^{2\,\mathrm{(I)}}$ as experimental input yields $Z=1.67\pm0.2$ 
-- unchanged in comparison with
the previous case ($Z$ is largely determined by $\Gamma_{a_{1}\rightarrow
\pi\gamma}$ which is the same in both $\chi^{2}$ calculations), 
$h_{1}=-116\pm70$, and $m_{\sigma}=(284\pm16)$ MeV. Note that in this case the
errors are much smaller than previously. The reason is that 
the mean value of $m_{\sigma}$ is almost on top of the two-pion decay
threshold and thus leads to an artificially small error band. 
For such small values of $m_\sigma$ the decay width $\Gamma_{\sigma
\rightarrow \pi\pi}$ is at least an order of magnitude smaller than the
physical value, but even for values of $m_\sigma$ up to 500 MeV (not
supported by our error analysis) the decay width never exceeds 150
MeV, see Fig.\ \ref{Sigmaf1}.

\subsection{Scenario II: Scalar Quarkonia Above 1 GeV}

\subsubsection{General discussion}

A possible way to resolve the problem of the unphysically small
two-pion decay width of the sigma meson is to identify the fields 
$\sigma$ and $a_{0}$ of the model with the
resonances $f_{0}(1370)$ and $a_{0}(1450)$, respectively. Thus, the scalar
quarkonium states are assigned to the energy region above 1 GeV. In the
following we investigate the consequences of this assignment. However, the
analysis cannot be conclusive for various reasons:
\begin{enumerate}
\item[(i)] 
The glueball field is missing. Many studies find that its role in the
mass region at about $1.5$ GeV is crucial, since it
mixes with the other scalar resonances.
\item[(ii)] The light scalar mesons below 1 GeV, such as $f_{0}(600)$ and
$a_{0}(980)$, are not included as elementary fields in our model. 
The question is if they can be dynamically generated 
from the pseudoscalar fields already present in our model by
solving a Bethe-Salpeter equation. If not, they should be
introduced as additional elementary fields from the very 
beginning [see also the discussion in Ref.\ \cite{dynrec}].
\item[(iii)] 
Due to absence of the resonance $f_{0}(600)$, the $\pi\pi$ scattering
length $a_0^0$ cannot be correctly described at tree-level: 
whereas $a_{0}^{2}$ stays always within the experimental error band, 
$a_{0}^{0}$ clearly requires a light scalar meson
for a proper description of experimental data because a
large value of $m_{\sigma}$ drives this quantity 
to the Weinberg limit ($\simeq 0.159$) which is outside the experimental
error band.
\end{enumerate} 
Despite these drawbacks, we turn to a quantitative analysis of this scenario.

\subsubsection{Decay of the $a_{0}(1450)$ meson}

As in Scenario I, the parameter $g_{2}$ can be expressed as a 
function of $Z$ by using the $\rho\rightarrow\pi\pi$ decay width 
(\ref{rhopionpion}). However, the parameter $h_{2}$
can no longer be fixed by the $f_1 \rightarrow a_0 \pi$ decay width:
the $a_0$ meson is now identified with the 
$a_{0}(1450)$ resonance listed in Ref.\ \cite{PDG}, with
a central mass of $m_{a_{0}}=1474$ MeV, and thus $f_1$ is too light to
decay into $a_0$ and $\pi$. One would be able to determine $h_2$
from the (energetically allowed) decay $a_0(1450) \rightarrow f_1 \pi$, 
but the corresponding decay width is not experimentally known.

Instead of performing a global fit, it is more convenient to proceed step by
step and calculate the parameters
$Z,\,h_{1},\,h_{2}$ explicitly. We vary $m_{\sigma}\equiv m_{f_{0}(1370)}$ 
within the experimentally known error band \cite{PDG} and
check if our result for $\Gamma_{f_{0}(1370)\rightarrow\pi\pi}$ 
is in agreement with experimental data.

We first determine $Z$ from $a_{1}\rightarrow\pi\gamma$, Eq.\
(\ref{a1piongamma}), and obtain $Z=1.67 \pm 0.21$. We then immediately
conclude that the $a_{1}\rightarrow\rho\pi$ decay width, Eq.\
(\ref{a1rhopion}), will remain the same as in Scenario I because this decay width
depends on $Z$ (which is virtually the same in both scenarios) 
and $g_{2}$ [which is fixed via
$\Gamma_{\rho\rightarrow\pi\pi}$, Eq.\ (\ref{g2Z}), in both scenarios].

The parameter $h_{1}$, being large-$N_{c}$ suppressed, will 
be set to zero in the present study.
We then only have to determine the parameter $h_{2}$. 
This is done by fitting the total decay
width of the $a_{0}(1450)$ meson to its experimental value \cite{PDG},
\begin{equation}
\Gamma_{a_{0}(1450)}(Z,h_{2})=\Gamma_{a_{0}\rightarrow\pi\eta}+\Gamma
_{a_{0}\rightarrow\pi\eta^{\prime}}+\Gamma_{a_{0}\rightarrow KK}+\Gamma
_{a_{0}\rightarrow\omega\pi\pi}\equiv\Gamma_{a_{0}(1450)}^{({\rm
exp})} =
(265\pm13)\text{ MeV.} \label{a01450}
\end{equation}

Although kaons have not been included into the calculations, we can easily
evaluate the decay into $KK$ by using flavour symmetry
\begin{eqnarray}
\Gamma_{a_{0}(1450)\rightarrow KK}(Z,h_{2})& =& 2\,
\frac{k(m_{a_{0}},m_{K},m_{K})}{8\pi m_{a_{0}}^{2}}
\left[ A_{a_{0}KK}(Z,h_{2}) \right]^{2}\; , \label{a0KK2} \\
A_{a_{0}KK}(Z,h_{2}) & =& 
\frac{1}{2 Z f_\pi}\left\{ m_{\eta_{N}}^{2}-m_{a_{0}}^{2}
+ \left( 1 - \frac{1}{Z^2} \right)  \left[  1-\frac{1}{2}
\frac{Z^{2}\phi^{2}}{m_{a_{1}}^{2}}(h_{2}-h_{3})\right]  
(m_{a_{0}}^{2}-2m_{K}^{2})  \right\}\; .
\end{eqnarray}

The remaining, experimentally poorly known decay width $\Gamma_{a_{0}
(1450)\rightarrow\omega\pi\pi}$ can be calculated from the sequential decay
$a_{0}\rightarrow\omega\rho\rightarrow\omega\pi\pi.$ Note that the
first decay step requires the $\rho$ to be slightly below its
mass-shell, since $m_{a_0} < m_\rho + m_\omega$. We denote
the off-shell mass of the $\rho$ meson by $x$. From the
Lagrangian (\ref{Lagrangian}) we obtain the following formula for the
$a_{0}\rightarrow\omega\rho$ decay width:
\[
\Gamma_{a_{0}(1450)\rightarrow\omega\rho}(x)
=\frac{k(m_{a_{0}},m_{\omega},x)}{8\pi m_{a_{0}}^{2}}
(h_{2}+h_{3})^{2}Z^{2}f_{\pi}^{2}\left[  3-\frac{x^{2}}{m_{\rho}^{2}}
+\frac{(m_{a_{0}}^{2}-x^{2}-m_{\omega}^{2})^{2}}{
4m_{\omega}^{2}m_{\rho}^{2}}\right]  \;.
\]
The full decay width $\Gamma_{a_{0}(1450)\rightarrow\omega\pi\pi
}$ is then obtained from the following equation:
\begin{equation}
\Gamma_{a_{0}(1450)\rightarrow\omega\pi\pi}=\int_0^\infty
\mathrm{d}x\, \Gamma_{a_{0}\rightarrow\omega\rho}(x)\, d_{\rho}(x)\;, 
\label{a0omegapionpion2}
\end{equation}
where $d_{\rho}(x)$ is the mass distribution of the $\rho$ meson, which is
taken to be of relativistic Breit-Wigner form:
\begin{equation}
d_{\rho}(x)    =N\, \frac{x^{2}\Gamma_{\rho\rightarrow\pi\pi}^{(\exp)}}
{(x^{2}-m_{\rho}^{2})^{2}+\left(x\Gamma_{\rho\rightarrow\pi\pi}^{(\exp)}
\right)^2}\,\theta(x-2m_{\pi})\;, \label{drho}
\end{equation}
where $\Gamma_{\rho\rightarrow\pi\pi}^{(\exp)}=149.1$ MeV and $m_{\rho}=775.49$
MeV \cite{PDG}. (In general, one should use the theoretical quantity
$\Gamma_{\rho\rightarrow\pi\pi}$, which is itself a function of $x$, instead
of $\Gamma_{\rho\rightarrow\pi\pi}^{(\exp)}$, see for instance Ref.\ \cite{lupo}
and refs.\ therein. 
This is, however, numerically irrelevant in the following.)
The normalization constant $N$ is chosen such that
\begin{equation}%
{\displaystyle\int\limits_{0}^{\infty}}
\mathrm{d}x\,d_{\rho}(x)=1\text{ ,}%
\end{equation}
in agreement with the interpretation of $\mathrm{d}x\,d_{\rho}(x)$ 
as the probability that the off-shell $\rho$ meson has a mass
between $x$ and $x+\mathrm{d}x.$

\begin{figure}
[h!]
\begin{center}
\includegraphics[
height=2.5988in,
width=3.9721in
]%
{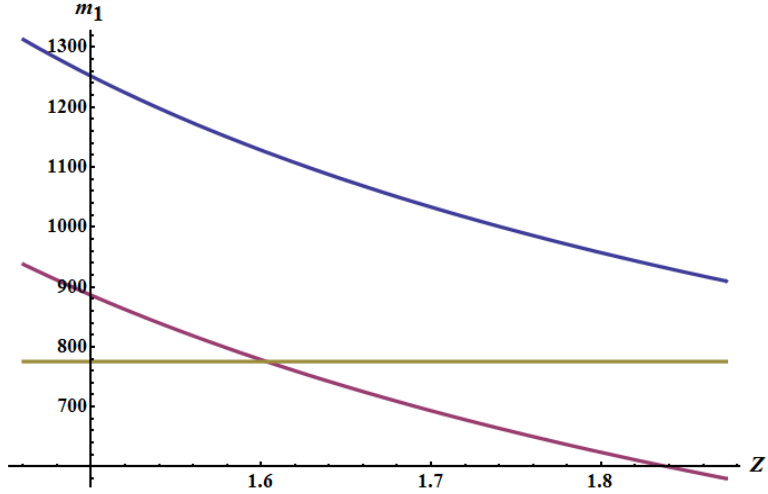}%
\caption{Dependence of $m_{1}$ on $Z$. The upper curve corresponds to the
lower band of $h_{2}$ values and the lower curve to the upper band of $h_{2}$
values. The horizontal line corresponds to the experimental value for
$m_{\rho}$.}%
\label{m1S2f}%
\end{center}
\end{figure}

Inserting Eqs.\ (\ref{a0etapion2}), (\ref{a0eta'pion2}), (\ref{a0KK2}), and
(\ref{a0omegapionpion2}) into Eq.\ (\ref{a01450}), we can express
$h_2$ as a function of $Z$, analogously to Eq.\ (\ref{g2Z}) where
$g_{2}$ was expressed as a function of $Z$. Similar to that case,
we obtain two bands for $h_2$, $-115\leq h_{2}\leq-20$ and $-25\leq h_{2}
\leq10$, the width of the bands corresponding to the uncertainty
in determining $Z$, $Z=1.67 \pm 0.21$. 
Both bands for $h_2$ remain practically unchanged if the $5\%$ experimental
uncertainty of $\Gamma_{a_{0}(1450)}^{({\rm exp})}$ is taken into account 
and thus we only use the mean value $265$ MeV in the following. 
Since $h_1$ is assumed to be zero, Eq.\ (\ref{m1eq}) allows to express
$m_1$ as a function of $Z$, $m_{1}=m_{1}(Z,h_{1}=0,h_{2}(Z))$ (we
neglect the experimental uncertainties of $m_\rho,\, m_{a_1}$, and
$f_\pi$). The result is shown in Fig.\ \ref{m1S2f}. 
The first band of (lower)
$h_{2}$ values should be discarded because it leads to
$m_{1}>m_{\rho}$. The second set of (higher) values leads to
$m_{1}<m_{\rho}$ only if the lower boundary for $Z$ is 1.60 rather
than 1.46. Thus, we shall use the set of larger $h_{2}$ values 
and take the constraint $m_{1}<m_{\rho}$ into account by
restricting the values for $Z$ to the range $Z=1.67_{-0.07}^{+0.21}$. 
As can be seen from Fig.\ \ref{m1S2f}, this sets a lower boundary
for the value of $m_1$, $m_1 \geq 580$ MeV.
Thus, in this scenario we obtain $m_{1}=720_{-140}^{+55}$ MeV. \newline

The values for the other parameters can be found in Table
\ref{Table2} (only 
central values are shown with the exception of $m_{1}$ where the
corresponding uncertainties are stated as well).%

%

\begin{table}[h!] \centering
\begin{tabular}
[c]{|c|c|c|c|c|c|c|c|c|c|c|c|}\hline
\textit{Parameter} & $h_{1}$ & $h_{2}$ & $h_{3}$ & $g_{1}$ & $g_{2}$ &
$m_{0}^{2}$ & $m_{1}$ & $\lambda_{1}$ & $\lambda_{2}$ & $c$ & $h_{0}$\\\hline
\textit{Value} & 0 & 4.7 & 2.4 & 6.4 & 3.1 & -811987 MeV$^{2}$ &
720$_{-140}^{+55}$ MeV & -3.6 & 84 & 88747 MeV$^{2}$ & $1\cdot10^{6}$
MeV$^{3}$\\\hline
\end{tabular}
\caption{Central values of the parameters for Scenario II.}\label{Table2}%
\end{table}%

$\newline$

Note that $\lambda_{1}\ll\lambda_{2},$ in agreement with the expectations from
the large-$N_{c}$ limit, Eq.\ (\ref{largen}). The value
of $m_{1}=720$ MeV is sizable and constitutes a dominant contribution
to the $\rho$ mass. This implies that non-quark contributions, for
instance a gluon condensate, play a
decisive role in the $\rho$ mass generation.

As a final step, we study the ratios $\Gamma_{a_{0}(1450)\rightarrow
\eta^{\prime}\pi}/\Gamma_{a_{0}(1450)\rightarrow\eta\pi}$ and
$\Gamma_{a_{0}(1450)\rightarrow K \overline{K}}/\Gamma_{a_{0}(1450)\rightarrow
\eta\pi}.$ Their experimental values read \cite{PDG}
\begin{equation}
\frac{\Gamma_{a_{0}(1450)\rightarrow\eta^{\prime}\pi}^{({\rm exp})}}{\Gamma
_{a_{0}(1450)\rightarrow\eta\pi}^{({\rm exp})}}=0.35\pm0.16\;\text{;}
\;\;\;\;\frac{\Gamma_{a_{0}(1450)\rightarrow K\overline{K}}^{({\rm exp})}}{
\Gamma_{a_{0}(1450)\rightarrow\eta
\pi}^{({\rm exp})}}=0.88\pm0.23\text{.}%
\end{equation}

Using the central value $Z=1.67$ and $\varphi = - 36^{\circ}$
for the $\eta-\eta'$ mixing angle, we obtain
$\Gamma_{a_{0}(1450)\rightarrow\eta^{\prime}\pi}/
\Gamma_{a_{0}(1450)\rightarrow \eta\pi}=1.0$ 
and $\Gamma_{a_{0}(1450)\rightarrow K\overline{K}}/\Gamma_{a_{0}(1450)
\rightarrow\eta\pi}=0.96$. The latter is in very good agreement
with the experiment, the former a factor of two larger. Note, however,
that according to Eqs.\ (\ref{a0etapion0}) and (\ref{a0etapion00}) 
the value of the ratio
$\Gamma_{a_{0}(1450)\rightarrow\eta^{\prime}\pi}/
\Gamma_{a_{0}(1450)\rightarrow \eta\pi}$ is proportional to 
$\sin^2 \varphi/\cos^2 \varphi$.
If a lower value of the angle is considered, e.g.,
$\varphi=-30^{\circ}$, then we obtain $\Gamma_{a_{0}(1450)\rightarrow
\eta^{\prime}\pi}/\Gamma_{a_{0}(1450)\rightarrow \eta\pi}=0.58$ for 
the central value of $Z$ and the central value of $\Gamma_{a_0(1450)}$
in Eq.\ (\ref{a01450}). Taking $Z=Z_{\max}$ and the upper boundary
$\Gamma_{a_{0}(1450)}^{({\rm exp})}=278$ MeV results in
$\Gamma_{a_{0}(1450)\rightarrow\eta^{\prime}\pi}/\Gamma_{a_{0}(1450)
\rightarrow\eta\pi}=0.48$, i.e., in agreement with the experimental
value. Therefore, our results in this scenario favour a smaller value 
of $\varphi$ than the one suggested by the KLOE Collaboration {\cite{KLOE}}.

It is possible to calculate the decay width $\Gamma_{a_{0}(1450)\rightarrow
\omega\pi\pi}$\ using Eq.\ (\ref{a0omegapionpion2}). We have obtained a very
small value $\Gamma_{a_{0}(1450)\rightarrow\omega\pi\pi}=0.1$ MeV. 
From Eq.\ (\ref{a0etapion2}) we obtain 
$\Gamma_{a_{0}(1450)\rightarrow\eta\pi}=$ 89.5
MeV, such that the ratio $\Gamma_{a_{0}(1450)\rightarrow\omega\pi\pi}/\Gamma
_{a_{0}(1450)\rightarrow\eta\pi}=0.0012$, in contrast to the results of
Ref.\ \cite{Baker}.

\subsubsection{Decay of the $f_{0}(1370)$ meson}

\begin{figure}
[h]
\begin{center}
\includegraphics[
height=2.514in,
width=5.3419in
]%
{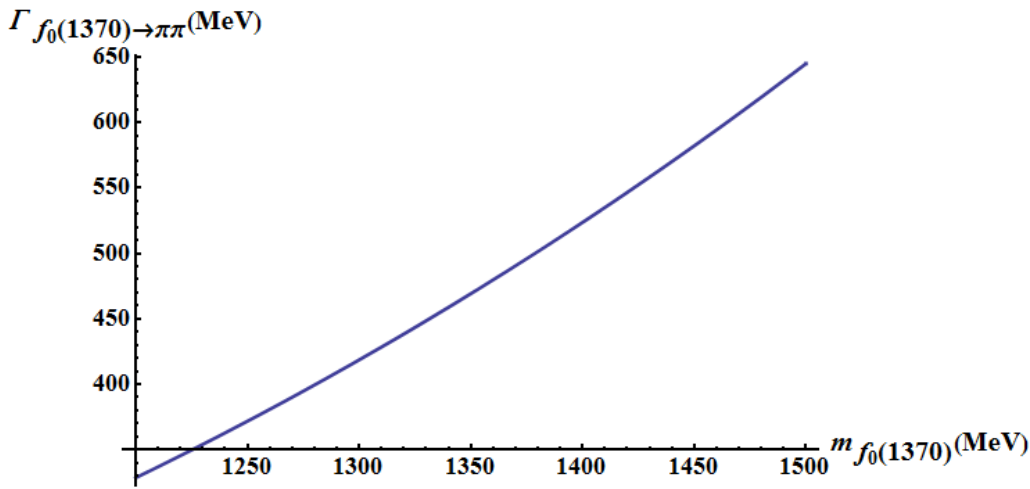}%
\caption{Dependence of the $f_{0}(1370)$ decay width on $m_{f_{0}(1370)}$. The
experimental value of the mass is expected to be in the range
1200--1500 MeV,
the width between 200 and 500 MeV \cite{PDG}.}%
\label{f0pionpionf}%
\end{center}
\end{figure}

It is now possible to calculate the width for the $f_{0}(1370)\rightarrow
\pi\pi$ decay using Eq.\ (\ref{sigmapionpion}). The decay width depends on the
$f_{0}(1370)$ mass, $Z$, $h_{1}$, and $h_{2}$ which is expressed via $Z$ using
Eq.\ (\ref{a01450}). The values of the latter three are listed 
in Table\textit{ }\ref{Table2}. In Fig.\ \ref{f0pionpionf} we show
the decay width as a function of the mass of $f_0(1370)$.

Assuming that the two-pion decay dominates the total decay width,
we observe a good agreement with the experimental values
if $m_{f_{0}(1370)}\alt 1380$ MeV. Other contributions to the decay
width are likely to reduce this upper bound on $m_{f_0(1370)}$ somewhat. 
Nevertheless, the
correspondence with the experiment is a lot better in this scenario where we
have identified $f_{0}(1370)$ rather than $f_{0}(600)$ as the 
(predominantly) isoscalar
$\bar{q}q$ state. Note that this result has been obtained using
the decay width of the $a_{0}(1450)$ meson (in order
to express $h_{2}$ via $Z$), which
is also assumed to be a scalar $\bar{q}q$ state in this scenario.

It is remarkable that vector mesons are crucial to obtain realistic
values for the decay width of $f_0(1370)$: without vector mesons, the
decay width is $\sim 10$ GeV and thus much too large.
This is why Scenario II has not been considered in
the standard linear sigma model. 

\bigskip%

The four-body decay $f_{0}(1370)\rightarrow 4\pi$ can also be studied.
Similarly to the $a_{0}(1450)\rightarrow\omega\rho$ decay, we view
$f_{0}(1370)\rightarrow4\pi$ as a sequential decay 
of the form $f_{0}(1370)\rightarrow
\rho\rho\rightarrow4\pi$. The Lagrangian (\ref{Lagrangian}) leads to%
\begin{equation}
\Gamma_{f_{0}(1370)\rightarrow\rho\rho}(x_{1},x_{2})=\frac{3}{16\pi}%
\frac{k(m_{f_{0}},x_{1},x_{2})}{m_{f_{0}}^{2}}\left(  h_{1}+h_{2}+h_{3}
\right)^{2}Z^{2}f_{\pi}^{2}\left[  4-\frac{x_{1}^{2}+x_{2}^{2}%
}{m_{\rho}^{2}}+\frac{(m_{f_{0}}^{2}-x_{1}^{2}-x_{2}^{2})^{2}}{4m_{\rho}^{4}%
}\right]  \text{,} \label{f0rhorho}%
\end{equation}
where $x_{1}$ and $x_{2}$ are the off-shell masses of the $\rho$
mesons. The decay width $\Gamma_{f_{0}\rightarrow4\pi}$ is then given by
\[
\Gamma_{f_{0}(1370)\rightarrow4\pi}=%
{\displaystyle\int\limits_{0}^{\infty}}
{\displaystyle\int\limits_{0}^{\infty}}
\mathrm{d}x_{1}\,\mathrm{d}x_{2}\,\Gamma_{f_{0}(1370)\rightarrow\rho\rho}%
(x_{1},x_{2})\,d_{\rho}(x_{1})\,d_{\rho}(x_{2})\;,
\]
with $\Gamma_{f_{0}(1370)\rightarrow\rho\rho}(x_{1},x_{2})$ from Eq.
(\ref{f0rhorho}) and $d_{\rho}(x)$ from Eq. (\ref{drho}).

Using the previous values for the parameters we obtain that the $\rho\rho$
contribution for the decay is small: $\Gamma_{f_{0}(1370)\rightarrow
\rho\rho\rightarrow4\pi}\simeq10 \pm 10$ MeV. (The error comes from
varying $Z$ between 1.6 and 1.88.) Ref.\ \cite{buggf0} quotes
54 MeV for the total $4 \pi$ decay width. Since Ref.\ \cite{abele}
ascertains that about 26\% of the total $4 \pi$ decay width originates
from the $\rho\rho$ decay channel, our result
is consistent with these findings.

\section{Conclusions and Outlook}

We have presented a linear sigma model with vector mesons and 
global chiral invariance. The motivation for considering 
global invariance rather than the standard Sakurai model with
local chiral invariance (with the exception of the vector-meson mass
term) was that the latter fails to describe
some important low-energy meson decay processes correctly, most notably the
two-pion decay width of the $\rho$ meson \cite{Lissabon}. This Lagrangian describes
mesons as pure quarkonium states. As shown in Sec.\ IV.A, the
resulting low-energy phenomenology is in general in good 
agreement with experimental data -- with one exception:
the model fails to correctly describe the $f_{0}(600)\rightarrow
\pi\pi$ decay width. This led us to conclude that $f_{0}(600)$
and $a_{0}(980)$ cannot be predominantly $\bar{q}q$ states. 

Assigning the scalar fields $\sigma$ and $a_{0}$ of the model to the
$f_{0}(1370)$ and $a_{0}(1450)$ resonances, respectively, 
improves the results for the decay widths considerably. We have
obtained $\Gamma_{f_{0}(1370)\rightarrow\pi\pi}\simeq300$-$500$ MeV for
$m_{f_{0}(1370)}=1200$-$1400$ MeV (see Fig.\ \ref{f0pionpionf}). Thus, the
scenario in which the scalar states above 1 GeV, $f_{0}(1370)$ and
$a_{0}(1450)$, are considered to be (predominantly) $\bar{q}q$ states appears
to be favoured over the assignment in which $f_{0}(600)$ and
$a_{0}(980)$ are considered (predominantly) $\bar{q}q$ states. 
However, a more detailed study of this
scenario is necessary, because a glueball state with the same
quantum numbers mixes with the quarkonium states. 
This allows to include the experimentally well-known
resonance $f_{0}(1500)$ into the study. 

Of course, interpreting $f_{0}(1370)$ and $a_{0}(1450)$ as $\bar{q}q$ states
leads to question about the nature of $f_{0}(600)$ and $a_{0}(980)$. Their
presence is necessary for the correct description of $\pi\pi$ scattering
lengths that differ from experiment for too large values of the isoscalar mass
(see Sec. IV.A.1). We distinguish two possibilities: (i) They can arise as
(quasi-)molecular states. This is possible if the attraction in the
$\pi\pi$ and $KK$ channels is large enough. In order to prove this,
one should solve the corresponding Bethe-Salpeter equation in the
framework of Scenario II.
In this case $f_{0}(600)$ and $a_{0}(980)$ can be classified as
genuinely dynamically generated states and should not appear in the
Lagrangian, see the discussion in Ref.\ \cite{dynrec}. If, however, the
attraction is not sufficient to generate the two resonances $f_{0}(600)$ and
$a_{0}(980)$ we are led to the alternative possibility that (ii) these two
scalar states must be incorporated into the model as additional tetraquark
states. In this case they shall appear from the very beginning in the
Lagrangian and should not be considered as dynamically generated states.
Of course, 
the isoscalar tetraquark, quarkonium, and glueball will mix to produce
$f_0(600),\, f_0(1370),$ and $f_0(1500)$, and the isovector
tetraquark and quarkonium will mix to produce $a_0(980)$ and $a_0(1450)$.

The issue of restoration of chiral symmetry at nonzero temperature and
density is one of the fundamental questions of modern hadron and
nuclear physics, see, e.g., Refs.\ \cite{RS,Heinz}. Linear sigma models
constitute an effective approach to study chiral
symmetry restoration because they
contain from the onset not only pseudoscalar and vector mesons, but also their
chiral partners with which they become degenerate once the chiral symmetry has
been restored. Once vacuum phenomenology is reasonably well
reproduced within our model,
we also plan to apply it to studies of chiral
symmetry restoration at nonzero temperatures and densities. 

Another important check of the model is the description
of the ALEPH data for the decay of the $\tau$ lepton into two and
three pions \cite{Anja}. In this way, we will have a better
constraint on the parameters of the model, e.g.\ the value for the
$a_{1}$ mass.

An extension of the model to $N_{f}=3$ can be performed \cite{PGR}; 
with the exception of the strange quark condensate, no further
free parameters will arise in this extension. However, 
much more data are available for the
strange mesons, which constitute an important test for the validity of
our approach.

\acknowledgments

The authors thank J.\ Schechter, F.\ Sannino, J.R.\ Pel\'{a}ez, S.\
Gallas, and S.\ Str\"{u}ber for valuable discussions. The work
of D.P.\ was partially supported by the Foundation ``Polytechnical
Society''. The work
of D.P.\ and F.G.\ was partially supported by BMBF. The work of D.H.R.\
was supported by the ExtreMe Matter Institute EMMI. This work was
(financially) supported by the Helmholtz International Center for FAIR
within the framework of the LOEWE program launched by the State of Hesse. 


\appendix

\section{The full Lagrangian}

This is the final form of the Lagrangian (\ref{Lagrangian}) that is obtained
after the shifts (\ref{shifts}) and the renormalisation of the pseudoscalar wave
functions; $\rho^{\mu\nu}\equiv\partial^{\mu}\rho^{\nu}%
-\partial^{\nu}\rho^{\mu}$; $a_{1}^{\mu\nu}\equiv\partial^{\mu}a_{1}^{\nu
}-\partial^{\nu}a_{1}^{\mu}$; ($\vec{A}$)$_{3}$ marks the third
component of the vector $\vec{A}$. Note that the term
$\mathcal{L}_{4}$ contains the (axial-)vector four-point vertices
[the terms $\sim g_{4,5,6,7}$ in the Lagrangian (\ref{Lagrangian})]. 
We do not give
the explicit form of $\mathcal{L}_{4}$ because it is not
relevant for the results that are presented in this paper.%

\begin{align*}
\lefteqn{\mathcal{L}=\frac{1}{2}\,(\partial^{\mu}\sigma+g_{1}Z\,\vec{\pi}%
\cdot\vec{a}_{1}^{\mu}+g_{1}wZ^{2}\,\partial^{\mu}\vec{\pi}\cdot\vec{\pi
}+g_{1}Z\eta f_{1}^{\mu}+g_{1}wZ^{2}\eta\,\partial^{\mu}\eta)^{2}}\\
&  -\,\frac{1}{2}\,\left[  m_{0}^{2}-c+3\left(  \lambda_{1}+\frac{\lambda_{2}%
}{2}\right)  \phi^{2}\right]  \sigma^{2}\\
&  +\,\frac{1}{2}\,(Z\partial^{\mu}\vec{\pi}+g_{1}Z\vec{\rho}^{\mu}\times
\vec{\pi}-g_{1}f_{1}^{\mu}\vec{a}_{0}-g_{1}wZ\partial^{\mu}\eta\vec{a}%
_{0}-g_{1}\sigma\vec{a}_{1}^{\mu}-g_{1}wZ\sigma\partial^{\mu}\vec{\pi})^{2}\\
&  +\,\frac{1}{2}\,(Z\partial^{\mu}\eta-g_{1}\sigma f_{1}^{\mu}-g_{1}%
wZ\sigma\partial^{\mu}\eta-g_{1}\,\vec{a}_{1}^{\mu}\cdot\vec{a}_{0}%
-g_{1}wZ\,\partial^{\mu}\vec{\pi}\cdot\vec{a}_{0})^{2}\\
&  -\,\frac{1}{2}\,\left[  m_{0}^{2}-c+\left(  \lambda_{1}+\frac{\lambda_{2}%
}{2}\right)  \phi^{2}\right]  Z^{2}\vec{\pi}^{2}-\frac{1}{2}\,\left[
m_{0}^{2}+c+\left(  \lambda_{1}+\frac{\lambda_{2}}{2}\right)  \phi^{2}\right]
Z^{2}\eta^{2}\\
&  +\,\frac{1}{2}\,[\partial^{\mu}\vec{a}_{0}+g_{1}\vec{\rho}^{\mu}\times
\vec{a}_{0}+g_{1}Zf_{1}^{\mu}\vec{\pi}+g_{1}wZ^{2}\vec{\pi}\partial^{\mu}%
\eta+g_{1}Z\eta\,\vec{a}_{1}^{\mu}+g_{1}wZ^{2}\eta\,\partial^{\mu}\vec{\pi
}]^{2}\\
&  -\,\frac{1}{2}\,\left[  m_{0}^{2}+c+\left(  \lambda_{1}+\frac{3}{2}%
\lambda_{2}\right)  \phi^{2}\right]  \vec{a}_{0}^{2}-\frac{\lambda_{2}}%
{2}[(\sigma\vec{a}_{0}+Z^{2}\eta\,\vec{\pi})^{2}+Z^{2}\vec{a}_{0}^{2}\vec{\pi
}^{2}-Z^{2}(\vec{a}_{0}\cdot\vec{\pi})^{2}]\\
&  -\,\frac{1}{4}\,\left(  \lambda_{1}+\frac{\lambda_{2}}{2}\right)
\,(\sigma^{2}+\vec{a}_{0}^{2}+Z^{2}\eta^{2}+Z^{2}\vec{\pi}^{2})^{2}-\left(
\lambda_{1}+\frac{\lambda_{2}}{2}\right)  \,\phi\sigma\,(\sigma^{2}+\vec
{a}_{0}^{2}+Z^{2}\eta^{2}+Z^{2}\vec{\pi}^{2})\\
&  -\,\lambda_{2}\phi\vec{a}_{0}\cdot(\sigma\vec{a}_{0}+Z^{2}\eta\,\vec{\pi
})-\frac{1}{4}\,(\partial^{\mu}\omega^{\nu}-\partial^{\nu}\omega^{\mu}%
)^{2}+\frac{m_{1}^{2}}{2}\,(\omega^{\mu})^{2}\\
&  -\,\frac{1}{4}\,[\partial^{\mu}\vec{\rho}^{\nu}-\partial^{\nu}\vec{\rho
}^{\mu}+g_{2}\vec{\rho}^{\mu}\times\vec{\rho}^{\nu}+g\,_{2}\vec{a}_{1}^{\mu
}\times\vec{a}_{1}^{\nu}+g_{2}wZ\partial^{\mu}\vec{\pi}\times\vec{a}_{1}^{\nu
}+g_{2}wZ\vec{a}_{1}^{\mu}\times\partial^{\nu}\vec{\pi}\\
&  +\,g_{2}w^{2}Z^{2}(\partial^{\mu}\vec{\pi}\ )\times(\partial^{\nu}\vec{\pi
})]^{2}\\
&  +\,\frac{m_{1}^{2}}{2}(\vec{\rho}^{\mu})^{2}-\frac{1}{4}\,(\partial^{\mu
}f_{1}^{\nu}-\partial^{\nu}f_{1}^{\mu})^{2}+\frac{m_{1}^{2}+g_{1}^{2}\phi^{2}%
}{2}\,(f_{1}^{\mu})^{2}+\frac{m_{1}^{2}+g_{1}^{2}\phi^{2}}{2}\,(\vec{a}%
_{1}^{\mu})^{2}\\
&  -\,\frac{1}{4}\,[\partial^{\mu}\vec{a}_{1}^{\nu}-\partial^{\nu}\vec{a}%
_{1}^{\mu}+g_{2}\vec{\rho}^{\mu}\times\vec{a}_{1}^{\nu}+g_{2}wZ\vec{\rho}%
^{\mu}\times\partial^{\nu}\vec{\pi}+g_{2}\vec{a}_{1}^{\mu}\times\vec{\rho
}^{\nu}+g_{2}wZ(\partial^{\mu}\vec{\pi})\times\vec{\rho}^{\nu}]^{2}\\
&  -\,g_{1}^{2}\,\phi\,\vec{a}_{1\mu}\cdot\lbrack\vec{\rho}^{\mu}\times
Z\vec{\pi}-f_{1}^{\mu}\vec{a}_{0}-wZ\vec{a}_{0}\,\partial^{\mu}\eta]\\
&  -\,g_{1}^{2}wZ\,\phi\,\partial_{\mu}\vec{\pi}\cdot\,[Z\vec{\rho}^{\mu
}\times\vec{\pi}-f_{1}^{\mu}\vec{a}_{0}-wZ\partial^{\mu}\eta\,\vec{a}%
_{0}]+g_{1}^{2}\,\phi f_{1\mu}\,(\vec{a}_{1}^{\mu}\cdot\vec{a}_{0}%
+wZ\partial^{\mu}\vec{\pi}\cdot\vec{a}_{0})\\
&  +\,g_{1}^{2}wZ\,\phi\,\partial_{\mu}\eta\,(\vec{a}_{1}^{\mu}\cdot\vec
{a}_{0}+wZ\partial^{\mu}\vec{\pi}\cdot\vec{a}_{0})\\
&  +\,g_{1}^{2}\,\phi\,\sigma\,[(f_{1}^{\mu})^{2}+2wZf_{1\,\mu}\partial^{\mu
}\eta+w^{2}Z^{2}(\partial^{\mu}\eta)^{2}]\\
&  +\,g_{1}^{2}\,\phi\,\sigma\,[(\vec{a}_{1}^{\mu})^{2}+2wZ\vec{a}_{1\mu}%
\cdot\partial^{\mu}\vec{\pi}+w^{2}Z^{2}(\partial^{\mu}\vec{\pi})^{2}]\\
&  -\,\frac{1}{2}\,\frac{g_{1}^{2}\phi^{2}}{\,m_{a_{1}}^{2}}\,Z^{2}%
\,(\partial_{\mu}\eta\partial^{\mu}\eta+\partial_{\mu}\vec{\pi}\cdot
\partial^{\mu}\vec{\pi})\\
&  +eA_{\mu}\{(\vec{a}_{0}\times\partial^{\mu}\vec{a}_{0})_{3}+Z^{2}(\vec{\pi
}\times\partial^{\mu}\vec{\pi})_{3}-4(\vec{\rho}^{\mu\nu}\times\vec{\rho}%
_{\nu})_{3}-4[(\vec{a}_{1\nu}+Zw\partial_{\nu}\vec{\pi})\times\vec{a}_{1}%
^{\mu\nu}]_{3}\\
&  +g_{1}\{2Z(f_{1}^{\mu}+Zw\partial^{\mu}\eta)(\vec{a}_{0}\times\vec{\pi
})_{3}+Z(\sigma+\phi)[(\vec{a}_{1}^{\mu}+Zw\partial^{\mu}\vec{\pi
})\times\vec{\pi}]_{3}\\
&  +\eta\lbrack\vec{a}_{0}\times(\vec{a}_{1}^{\mu}+Zw\partial^{\mu
}\vec{\pi})]_{3}-a_{0}^{3}(a_{0}^{1}\rho^{\mu1}+a_{0}^{2}\rho^{\mu2})-Z^{2}%
\pi^{3}(\pi^{1}\rho^{\mu1}+\pi^{2}\rho^{\mu2})\\
&  +\rho^{\mu3}[(a_{0}^{1})^{2}+(a_{0}^{2})^{2}+Z^{2}(\pi^{1})^{2}+Z^{2}%
(\pi^{2})^{2}]\}\\
&  +4g_{2}\{[\vec{\rho}_{\nu}^{2}+\vec{a}_{1\nu}^{2}+Z^{2}w^{2}\left(
\partial_{\nu}\vec{\pi}\right)^{2}+Zw\vec{a}_{1\nu}\cdot\partial^{\nu}%
\vec{\pi}]\rho^{\mu3}+2\vec{\rho}_{\nu}\cdot(\vec{a}_{1}^{\nu}+Zw\partial
^{\nu}\vec{\pi})\times(a_{1}^{\mu3}+Zw\partial^{\mu}\pi^{3})\\
&  -(\vec{\rho}_{\nu}\cdot\vec{\rho}^{\mu}+\vec{a}_{1\nu}\cdot\vec{a}_{1}%
^{\mu}+Zw\vec{a}_{1\nu}\cdot\partial^{\mu}\vec{\pi}+Zw\vec{a}_{1}^{\mu}%
\cdot\partial_{\nu}\vec{\pi}+Z^{2}w^{2}\partial_{\nu}\vec{\pi}\cdot
\partial^{\mu}\vec{\pi})\rho^{\nu3}\\
&  -(\vec{\rho}^{\mu}\cdot\vec{a}_{1\nu}+\vec{a}_{1}^{\mu}\cdot\vec{\rho}%
_{\nu}+Zw\vec{\rho}^{\mu}\cdot\partial_{\nu}\vec{\pi}+Zw\vec{\rho}_{\nu}%
\cdot\partial^{\mu}\vec{\pi})a_{1}^{\nu3}\}\}\\
&  +\frac{e^{2}}{2}A_{\mu}A^{\mu}[(a_{0}^{1})^{2}+(a_{0}^{2})^{2}+Z^{2}%
(\pi^{1})^{2}+Z^{2}(\pi^{2})^{2}+4(\rho^{\nu1})^{2}+4(\rho^{\nu2})^{2}\\
&  +4(a_{1\nu}^{1}+Zw\partial_{\nu}\pi^{1})^{2}+4(a_{1\nu}^{2}+Zw\partial
_{\nu}\pi^{2})^{2}]\\
&  -2e^{2}A_{\mu}A_{\nu}[\rho^{\mu1}\rho^{\nu1}+\rho^{\mu2}\rho^{\nu2}%
+(a_{1}^{\mu1}+Zw\partial^{\mu}\pi^{1})(a_{1}^{\nu1}+Zw\partial^{\nu}\pi
^{1})\\
&  +(a_{1}^{\mu2}+Zw\partial^{\mu}\pi^{2})(a_{1}^{\nu2}+Zw\partial^{\nu}%
\pi^{2})]+{\mathcal{L}}_{h_{1,2,3}}+\mathcal{L}_{g_{3}}+\mathcal{L}_{4}%
\end{align*}

\begin{align*}
{\mathcal{L}}_{h_{1,2,3}}  &  =\left(  \frac{h_{1}}{4}+\frac{h_{2}}{4}%
+\frac{h_{3}}{4}\right)  \left(  \sigma^{2}+2\phi\sigma+Z^{2}\eta^{2}+\vec
{a}_{0}^{2}+Z^{2}\vec{\pi}^{2}\right)  (\omega_{\mu}^{2}+\vec{\rho}_{\mu}%
^{2})\\
&  +\left(  \frac{h_{1}}{4}+\frac{h_{2}}{4}-\frac{h_{3}}{4}\right)  \left(
\sigma^{2}+2\phi\sigma+Z^{2}\eta^{2}+\vec{a}_{0}^{2}+Z^{2}\vec{\pi}%
^{2}\right)  \{f_{1\mu}^{2}+\vec{a}_{1\mu}^{2}+Z^{2}w^{2}[\left(
\partial_{\mu}\eta\right)  ^{2}+\left(  \partial_{\mu}\vec{\pi}\right)
^{2}]\\
&  +2Zw(f_{1\mu}\partial^{\mu}\eta+\vec{a}_{1\mu}\cdot\partial^{\mu}\vec{\pi
})\}\\
&  +\left(  \frac{h_{1}}{4}+\frac{h_{2}}{4}+\frac{h_{3}}{4}\right)  \phi
^{2}(\omega_{\mu}^{2}+\vec{\rho}_{\mu}^{2})+\left(  \frac{h_{1}}{4}%
+\frac{h_{2}}{4}-\frac{h_{3}}{4}\right)  \phi^{2}(f_{1\mu}^{2}+\vec{a}_{1\mu
}^{2})\\
&  +(h_{2}+h_{3})\omega_{\mu}[(\sigma+\phi)\vec{a}_{0}+Z^{2}\eta\vec{\pi
}]\cdot\vec{\rho}^{\mu}\\
&  +(h_{2}-h_{3})[(\sigma+\phi)\vec{a}_{0}+Z^{2}\eta\vec{\pi}]\cdot\lbrack
f_{1\mu}\vec{a}_{1}^{\mu}+Zw(\vec{a}_{1\mu}\partial^{\mu}\eta+f_{1\mu}%
\partial^{\mu}\vec{\pi})+Z^{2}w^{2}(\partial_{\mu}\eta)(\partial^{\mu}\vec
{\pi})]\\
&  +(h_{2}+h_{3})Z(\vec{a}_{0}\times\vec{\pi})\cdot(\omega_{\mu}\vec{a}%
_{1}^{\mu}+Zw\omega_{\mu}\partial^{\mu}\vec{\pi})\\
&  +(h_{2}-h_{3})Z(\vec{a}_{0}\times\vec{\pi})\cdot(f_{1\mu}\vec{\rho}^{\mu
}+Zw\vec{\rho}_{\mu}\partial^{\mu}\eta)\\
&  +h_{3}Z[\eta\vec{a}_{0}-(\sigma+\phi)\vec{\pi}]\cdot\lbrack\vec{\rho}_{\mu
}\times(\vec{a}_{1}^{\mu}+Zw\partial^{\mu}\vec{\pi})]\\
&  -\frac{h_{3}}{2}\{(\vec{a}_{0}\times\vec{\rho}^{\mu})^{2}-[\vec{a}%
_{0}\times(\vec{a}_{1}^{\mu}+Zw\partial^{\mu}\vec{\pi})]^{2}+Z^{2}(\vec{\pi
}\times\vec{\rho}^{\mu})^{2}-Z^{2}[\vec{\pi}\times(\vec{a}_{1}^{\mu
}+Zw\partial^{\mu}\vec{\pi})]^{2}\}
\end{align*}

\begin{align*}
{\mathcal{L}}_{g_{3}}  &  =-4g_{3}\{\partial_{\mu}\omega_{\nu}[\omega^{\mu
}\omega^{\nu}+f_{1}^{\mu}f_{1}^{\nu}+\vec{\rho}^{\mu}\cdot\vec{\rho}^{\nu
}+\vec{a}_{1}^{\mu}\cdot\vec{a}_{1}^{\nu}+Zw(f_{1}^{\mu}\partial^{\nu}%
\eta+f_{1}^{\nu}\partial^{\mu}\eta+\vec{a}_{1}^{\mu}\cdot\partial^{\nu}%
\vec{\pi}+\vec{a}_{1}^{\nu}\cdot\partial^{\mu}\vec{\pi})\\
&  +Z^{2}w^{2}(\partial^{\mu}\eta\partial^{\nu}\eta+\partial^{\mu}\vec{\pi
}\cdot\partial^{\nu}\vec{\pi})]\\
&  +(\partial_{\mu}f_{1\nu}+Zw\partial_{\mu}\partial_{\nu}\eta)[\omega^{\mu
}f_{1}^{\nu}+\omega^{\nu}f_{1}^{\mu}+\vec{\rho}^{\mu}\cdot\vec{a}_{1}^{\nu
}+\vec{\rho}^{\nu}\cdot\vec{a}_{1}^{\mu}+Zw(\omega^{\mu}\partial^{\nu}%
\eta+\omega^{\nu}\partial^{\mu}\eta+\vec{\rho}^{\mu}\cdot\partial^{\nu}%
\vec{\pi}+\vec{\rho}^{\nu}\cdot\partial^{\mu}\vec{\pi})]\\
&  +\partial_{\mu}\vec{\rho}_{\nu}\cdot\lbrack\omega^{\mu}\vec{\rho}^{\nu
}+\omega^{\nu}\vec{\rho}^{\mu}+f_{1}^{\mu}\vec{a}_{1}^{\nu}+f_{1}^{\nu}\vec
{a}_{1}^{\mu}+Zw(\vec{a}_{1}^{\mu}\partial^{\nu}\eta+\vec{a}_{1}^{\nu}%
\partial^{\mu}\eta+f_{1}^{\mu}\partial^{\nu}\vec{\pi}+f_{1}^{\nu}\partial
^{\mu}\vec{\pi})+Z^{2}w^{2}(\partial^{\mu}\eta\partial^{\nu}\vec{\pi}%
+\partial^{\nu}\eta\partial^{\mu}\vec{\pi})]\\
&  +(\partial_{\mu}\vec{a}_{1\nu}+Zw\partial_{\mu}\partial_{\nu}\vec{\pi
})\cdot\lbrack f_{1}^{\mu}\vec{\rho}^{\nu}+f_{1}^{\nu}\vec{\rho}^{\mu}%
+\omega^{\mu}\vec{a}_{1}^{\nu}+\omega^{\nu}\vec{a}_{1}^{\mu}+Zw(\vec{\rho
}^{\mu}\partial^{\nu}\eta+\vec{\rho}^{\nu}\partial^{\mu}\eta+\omega^{\mu
}\partial^{\nu}\vec{\pi}+\omega^{\nu}\partial^{\mu}\vec{\pi})]\}\\
&  +4eg_{3}A_{\mu}\{\omega_{\nu}[(\vec{\rho}^{\mu}\times\vec{\rho}^{\nu}%
)_{3}+(\vec{a}_{1}^{\mu}\times\vec{a}_{1}^{\nu})_{3}+Zw(\partial^{\mu}\vec
{\pi}\times\vec{a}_{1}^{\nu})_{3}+Zw(\vec{a}_{1}^{\mu}\times\partial^{\nu}%
\vec{\pi})_{3}+Z^{2}w^{2}(\partial^{\mu}\vec{\pi}\times\partial^{\nu}\vec{\pi
})_{3}]\\
&  +(f_{1\nu}+Zw\partial_{\nu}\eta)[(\vec{\rho}^{\mu}\times\vec{a}_{1}^{\nu
})_{3}+(\vec{a}_{1}^{\mu}\times\vec{\rho}^{\nu})_{3}+Zw(\vec{\rho}^{\mu}%
\times\partial^{\nu}\vec{\pi})_{3}+Zw(\partial^{\mu}\vec{\pi}\times\vec{\rho
}^{\nu})_{3}]\}
\end{align*}

\end{document}